\documentclass[12pt]{article}

\usepackage{amsmath}
\usepackage{amssymb}
\usepackage{amsfonts}
\usepackage{latexsym}
\usepackage{braket}
\usepackage{color}
\usepackage{graphicx}
\usepackage{physics}
\usepackage{subcaption}
\usepackage{hyperref}
\catcode `\@=11 \@addtoreset{equation}{section}
\def\theequation{\arabic{section}.\arabic{equation}}
\catcode `\@=12

\newcommand{\be}{\begin{equation}}
	\newcommand{\en}{\end{equation}}
\newcommand{\bea}{\begin{eqnarray}}
	\newcommand{\ena}{\end{eqnarray}}
\newcommand{\beano}{\begin{eqnarray*}}
	\newcommand{\enano}{\end{eqnarray*}}
\newcommand{\bee}{\begin{enumerate}}
	\newcommand{\ene}{\end{enumerate}}

\newcommand{\R}{\mathbb{R}}
\newcommand{\mc}{\mathcal}

\newcommand{\Sc}{{\cal S}}

\newcommand{\F}{{\cal F}}
\newcommand{\G}{{\cal G}}

\newcommand{\Lc}{{\cal L}}

\newcommand{\1}{1 \!\! 1}

\newcommand{\ltwo}{{\Lc^2(\mathbb{R})}}
\newcommand{\scr}{{\Sc(\mathbb{R})}}

\newcommand{\Hil}{\mc H}

\renewcommand{\l}{\langle}
\renewcommand{\r}{\rangle}
\newcommand{\pin}[2]{\l#1 , #2\r}

\newcommand{\arcsinh}{\,\text{arcsinh}}

\catcode `\@=11 \@addtoreset{equation}{section}
\catcode `\@=12

\begin{document}
	
	\thispagestyle{empty}

	\vspace*{2cm}
	
	\begin{center}
		{\Large \bf A position dependent mass Hamiltonian and abstract ladder operators}   
		
		\vspace{5mm}
		
		{\large F. Bagarello\footnote{Corresponding author}}\\
		Dipartimento di Ingegneria,
		Universit\`a di Palermo,\\ I-90128  Palermo, Italy\\
		and I.N.F.N., Sezione di Catania\\
		e-mail: fabio.bagarello@unipa.it\\

		\vspace{8mm}
		
		{\large E. Balistreri}\\
		Dipartimento di Fisica e Chimica,
		Universit\`a di Palermo,\\ I-90128  Palermo, Italy\\
		e-mail: emanuele.balistreri@community.unipa.it\\

		\vspace{8mm}
		
		{\large A. Faddetta}\\
		Dipartimento di Scienze Matematiche e Informatiche, Scienze Fisiche e Scienze della Terra,  Universit\`a di Messina,
		Viale F. Stagno d'Alcontres 31, 
		I--98166 Messina, Italy\\
		e-mail: antonino.faddetta@studenti.unime.it\\

	\end{center}
	
	\vspace*{1cm}
	
	\begin{abstract}
		\noindent We consider the Hamiltonian $H$ of a particle in one dimension with a position dependent mass for which we apply the recent strategy of the so-called {\em abstract ladder operators}, in the attempt to find its eigenvalues and eigenvectors. We don't assume that $H$ is self-adjoint, while we focus on the case of a factorizable operator. We show then that pseudo-bosonic operators play a relevant role in this analysis, and we construct bi-coherent states attached to these operators. Explicit examples are discussed.
	\end{abstract}
	
	\vspace{2cm}
	
	
	\vfill

	
	\newpage
	
	\section{Introduction}
	
	One of the essential steps in the analysis of many physical systems $\Sc$ consists in writing first the Hamiltonian $H$ for $\Sc$, and then in diagonalizing it, i.e. in finding its eigenvalues and eigenvectors. Since the beginning of Quantum Mechanics, for several decades, the various observables of $\Sc$ were assumed to be self-adjoint operators in order to have only real eigenvalues and unitary time evolution. However, after the seminal paper \cite{benboet}, it started to be clear that this assumption is indeed too strong, and not quite necessary, other than being more mathematically than physically motivated. This is particularly relevant in connection with open systems, where part of the energy of a (smaller) system is exchanged with its (larger) environment. Since the appearance of \cite{benboet} a lot of interest arose in dealing with manifestly non self-adjoint Hamiltonians, and related stuff. The mathematics which is behind these operators is quite interesting and non trivial, and the physical consequences are also quite interesting. We refer to \cite{specissue2012}-\cite{mosta} for some books, reviews, and edited volumes on these topics, where a lot of other references can be found. 
	
	In recent years many researchers started to consider  systems with a position dependent mass, see \cite{pdmass1}-\cite{quesne} and references therein.  This interest gave rise to several models, and many different aspects of these systems have been, and still are, discussed, also at a relativistic level, see \cite{art2023,art2006} for example. Soon after their appearance, also extentions to non self-adjoint Hamiltonians have been proposed, as in \cite{art2008a,art2008b,art2025}. Our paper originates from exactly from this interest: we look for the eigenvalues and the eigenvectors of a certain (non self-adjoint) Hamiltonian with position dependent mass. But, with respect to what is done in the cited papers, we adopt here a method recently proposed by one of us and which refers to the so-called {\em abstract ladder operators} (ALOs), \cite{fern1}-\cite{bagalo2}.  Following this idea, and requiring further that $H$ can be factorized, we find that this is possible indeed, and that the operators needed to factorize $H$ are pseudo-bosonic operators, \cite{bagspringerbook}. Adopting  then a rather general construction, we diagonalize $H$, finding also conditions which guarantee that the eigenvectors of $H$ and $H^\dagger$ are square-integrable. This is the content of Section \ref{sect2}. In Section \ref{sect3} we construct the bi-coherent states connected to the pseudo-bosonic lowering operators, \cite{bagspringerbook}. In Section \ref{sect4}  the same Hamiltonian is considered from a different point of view, relating $H$ with the Hamiltonian of a quantum harmonic oscillator by introducing an invertible and unbounded transformation. We will recover, with a different technique, the same results, i.e. the same eigenvectors and eigenvalues obtained before in this paper. In Section \ref{sect5} we propose three different examples corresponding to three different choices of mass $m(x)$, describing three different cases met previously in the general analysis of $H$. Our conclusions are given in Section \ref{sect6}. In the Appendices we review some useful results on ALOs, and we give the details of some explicit computations.

	\section{The Hamiltonian and its ALO}\label{sect2}

	The Hamiltonian we are interested in is 
	\begin{equation}
		H=-\frac{\hbar^2}{2m(x)}\dv[2]{}{x}+\frac{\hbar^2m'(x)}{2m^2(x)}\dv{}{x}+V(x),
		\label{21}\end{equation}
	recently considered in \cite{estrada}, but widely studied by several authors. 
	We only cite here  \cite{takou,quesne}, where many other related references can be found. With respect with what is usually assumed in the literature, we relax the condition of having a real potential $V(x)$. In fact, we write  $V(x)=V_R(x)+iV_I(x)$, and we do not require $V_I(x)$ to be zero. As for $m(x)$, for the moment we only require that $m(x)>0$, as any {\em physical} mass should be, and that it is differentiable. In this way $H$ is, at least formally, well defined.
	
	Our aim is to discuss if and when the Hamiltonian (\ref{21})  can be diagonalized using the idea first proposed in \cite{fern1,fern2,fern3}, and then extended in \cite{bagalo1,bagalo2}. We refer to the Appendix A for a short review of the results which are useful for us here.
	
	Our first aim is to find an operator $A$ which obeys the commutation rule in (\ref{a1}) for $N=1$:
	\be
	[H,A]=\lambda A,
	\label{22}\en
	for some fixed $\lambda\in\mathbb{C}$.  In what follows we will only consider $\lambda\neq0$, to avoid the case in which $H$ and $A$ simply commute. In this paper we assume that $A$ has the following expression:
	\be
	A=\frac{1}{\sqrt{2}}\left(\alpha(x)\dv{}{x}+\beta(x)\right),
	\label{23}\en
	where we allow for complex expressions of $\alpha(x)=\alpha_R(x)+i\alpha_I(x)$ and $\beta(x)=\beta_R(x)+i\beta_I(x)$. 
	
	This class of operators have been considered recently by one of us, see \cite{bagspringerbook} and references therein.
	
	From now on, except when explicitly stated, we assume that $\alpha(x), \beta(x)$ and $m(x)$ are $C^\infty$ functions. Moreover, we also assume for the moment that $\frac{1}{m(x)}$ and $\frac{m'(x)}{m^2(x)}$, together with $\alpha(x)$ and $\beta(x)$,  grow at most polynomially for $|x|$ diverging. Under these assumptions both $H$ and $A$ leave $\scr$, the Schwartz space, stable: if $f\in\scr$ then $Hf,Af\in\scr$. These conditions are not really essential, and can be relaxed. This is in fact what happens in some explicit examples considered in Section \ref{sect5}. 
	
	\vspace{2mm}
	
	{\bf Remark:--} What we have just observed implies, in particular, that both $H$ and $A$ are densely defined. This is a crucial property  unbounded operators must satisfy, to allow to carry on rigorous computations, \cite{reed}.
	
	\vspace{2mm}
	
		It is clear that the  potential $V(x)$ in equation \eqref{21} cannot be arbitrary. Indeed its expression is naturally restricted by the request in \eqref{22} and by the assumption that $A$ can be written as in \eqref{23}. This will produce the expression (\ref{27}) below.
	
	Condition (\ref{22}) is well defined (for instance) on $\scr$ and it is satisfied if the following system holds:
	
	\begin{equation}
		\label{24}
		\begin{cases}
			
			\quad \alpha'(x) + \frac{ m'(x)}{2m(x)}\alpha(x) = 0, \\[10pt]
			
			\quad 
			\beta'(x)=
			\alpha(x)\left(  \frac{3(m'(x))^2}{8m^2(x)}-\frac{m''(x)}{4m(x)}-\frac{\lambda m(x)}{\hbar^2} \right), \\[10pt]
			
			\quad V'(x)=\frac{\hbar^2}{32}\left(\frac{4m'''(x)}{m^2(x)}+\frac{21(m'(x))^3}{m^4(x)}-\frac{22m''(x)m'(x)}{m^3(x)} \right)-\lambda\left(\frac{\beta(x)\overline{\alpha(x)}}{\abs{\alpha(x)}^2}+\frac{m'(x)}{4m(x)}\right).
		\end{cases}
	\end{equation}
	The first equation can be easily solved:
	\begin{equation}
		\label{25}
		\alpha(x)=\frac{\delta}{\sqrt{m(x)}}
	\end{equation}
	where $\delta\in\mathbb{C}$ is a complex integration constant. From the second equation, we can easily find that 
	\begin{equation}
		\label{26}
		\beta(x)=\delta\left[-\frac{m'(x)}{4m^{3/2}(x)}-\frac{\lambda}{\hbar^2}F(x)+\gamma\right], \quad F(x)=\int \sqrt{m(x)}dx
	\end{equation}
	where, again, $\gamma \in \mathbb{C}$. We observe that $\beta(x)$ indeed depends on $\lambda$ which, at this stage, is not fixed. From (\ref{23}) we see that $\delta$ behaves as an overall multiplicative parameter in $A$. Due to (\ref{22}) we easily understand that its role is not really relevant, and for this reason   we set $\delta=1$ in what follows.
	Replacing now $\alpha(x)$ and $\beta(x)$ in $(\ref{24})_3$ we find that
	\begin{equation}
		\label{27}
		V(x)=\frac{\hbar^2}{8}\left(\frac{m''(x)}{m^2(x)}-\frac{7(m'(x))^2}{4m^3(x)}\right)+\frac{\lambda^2}{2\hbar^2}F^2(x)-\lambda\gamma F(x)
	\end{equation}
	where   we have neglected a constant of integration since it only affects $H$ in (\ref{21}) for an unessential additive constant. We observe that, like $\beta(x)$, also $V(x)$ depends on $\lambda$.

	\vspace{2mm}
	
	{\bf Remark:--} It is maybe worth spending few lines on the role of $\lambda$. Here we are considering $\lambda$ as a free parameter: different $\lambda$'s produce different $H$ and $A$. Some of them could have a physical meaning, some others might not. For instance, in what follows we will show that, in rather general conditions, $\lambda$ must have a non positive real part, if we want to work with square integrable functions. Hence not all $\lambda$'s are allowed.  Furthermore, it might happen that in some specific situations the explicit value of $\lambda$ is simply fixed by $H$ and $A$, rather than the other way around. This is what happens for the harmonic oscillator where (in suitable units) we have $H_{ho}=a^\dagger a+\frac{1}{2}\1$, with $[a,a^\dagger]=\1$. Now, if we take $A=a$, we deduce that $[H_{ho},A]=-A$, so that $\lambda=-1$.  But, since we also have $[H_{ho},a^\dagger]=a^\dagger$, we can rather identify $A$ with $a^\dagger$ and, once again, $\lambda$ is fixed: $\lambda=1$. These two choices, however, produce rather different consequences: $\lambda=-1$ gives rise to square-integrable functions, while $\lambda=1$ doesn't. We will return on the role of $\lambda$, especially in connection with square-integrability of the eigenvectors of $H$,  later in this section. 
	
	\vspace{1mm}

	

	Summarizing, for this moment, we work with the following $H$ and $A$:
	\be
	H=-\frac{\hbar^2}{2m(x)}\dv[2]{}{x}+\frac{\hbar^2m'(x)}{2m^2(x)}\dv{}{x}+\frac{\hbar^2}{8}\left(\frac{m''(x)}{m^2(x)}-\frac{7(m'(x))^2}{4m^3(x)}\right)+\frac{\lambda^2}{2\hbar^2}F^2(x)-\gamma\lambda F(x),
	\label{28}\en
	and
	\be
	A =\frac{1}{\sqrt{2}}\left[\frac{1}{\sqrt{m(x)}}\dv{}{x}-\frac{ m'(x)}{4m^{3/2}(x)}-\frac{\lambda}{\hbar^2}F(x)+ \gamma\right].
	\label{29}\en
	
	It may be useful to remark that, if we restrict our analysis to a real potential $V(x)$, this is indeed the case if  $\lambda$ and $\gamma$ are both real. Hence all the formulas we have deduced here  collapse to those in \cite{estrada}. It is also useful to stress once more that formulas (\ref{28}) and (\ref{29}) clearly show that the particular value of $\lambda$ in (\ref{22}) fixes a specific expression of $H$ and $A$. In other words: different $\lambda$'s correspond to different operators $H$ and $A$. Notice that, at this stage, we have not found any specific constraint on $\lambda$. Each $\lambda$, if $H$ and $A$ are chosen as in (\ref{28}) and (\ref{29}), is such that condition (\ref{22}) is satisfied. 
	
	The vacuum of $A$, $A\varphi_0(x)=0$, can be easily computed by solving the differential equation
	$$
	\frac{1}{\sqrt{2}}\left[\frac{1}{\sqrt{m(x)}}\dv{}{x}-\frac{ m'(x)}{4m^{3/2}(x)}-\frac{\lambda}{\hbar^2}F(x)+\gamma\right]\varphi_0(x)=0,
	$$
	whose solution is
	\begin{equation}
		\varphi_0(x)=N_{\varphi,0}m^{\frac{1}{4}}(x)e^{\frac{\lambda}{2\hbar^2}F^2(x)-\gamma F(x)},
		\label{210}
	\end{equation}
	where $N_{\varphi,0}$ is an integration constant. We will show in (\ref{211}) that $\varphi_0(x)$, other than satisfying $A\varphi_0(x)=0$, is an eigenstate of $H$ with eigenvalue $E_0=-\frac{1}{2}\left(\gamma^2\hbar^2+\lambda\right)$.  In order for $\varphi_0(x)$ to be square integrable it is sufficient to have $\Re\left(\frac{\lambda}{2\hbar^2}F^2(x)-\gamma F(x)\right)<0$ and $\Re\left(\frac{\lambda}{2\hbar^2}F^2(x)-\gamma F(x)\right)\rightarrow-\infty$ for $|x|\rightarrow\infty$.
	
	\subsection{The simplest case: the harmonic oscillator}
	
	To understand the meaning of these conditions, and if they are reasonable or not, we go back to $H$ in (\ref{21}) and we put $m(x)=\hbar^2$. With this choice $H$ simplifies: $H=-\frac{1}{2}\dv[2]{}{x}+V(x)$ where, see (\ref{210b}) below, $V(x)$ turns out to be quadratic in $x$. Hence we are going back to an harmonic oscillator (shifted or not, self-adjoint or not. This depends on our choice of the other parameters, like $\lambda$ for instance). From (\ref{25}), (\ref{26}) and (\ref{27}) we get the following:
	\be
	\alpha(x)=\frac{1}{\hbar},\qquad F(x)=\hbar x, \qquad \beta(x)=-\frac{\lambda}{\hbar}x+\gamma, \qquad V(x)=\frac{\lambda^2}{2}x^2-\lambda\gamma\hbar x,
	\label{210b}\en
	and
	$$
	A=\frac{1}{\sqrt{2}\hbar}\left(\dv{}{x}-\lambda\,x+ \hbar\gamma\right),
	$$
	where we have also used the fact that $\delta=1$. A straightforward check shows that in this case $[H,A]=\lambda A$ is satisfied for all possible $\lambda$. It is clear from (\ref{210b}) that, in order for $V(x)$ to be real, so to  have a (at least formally) self-adjoint Hamiltonian $H$, both $\lambda$ and $\gamma$ must be real, or at least such that $\lambda^2$ and $\lambda\gamma$ are real. A possible solution of this latter situation is  $\lambda=\gamma=i$. In this case, we recover a shifted version of the inverted harmonic oscillator, see \cite{barton}-\cite{bagbaku} and references therein.
	
	Going back to our original problem, and only focusing on the nature of $\varphi_0(x)$, since 
	$$
	\Re\left(\frac{\lambda}{2\hbar^2}F^2(x)-\gamma F(x)\right)=\Re\left(\frac{\lambda}{2}x^2-\gamma \hbar x\right),
	$$
	we conclude that $\varphi_0(x)\in\ltwo$ only if $\Re(\lambda)<0$.  We have already observed that this is exactly what happens for  $H_{ho}=a^\dagger a+\frac{1}{2}\1$, when identifying $A$ with $a$, $[a,a^\dagger]=\1$: $[H_{ho},a]=-a$, so that $\lambda=-1$, while it is not the case if we rather identify $A$ with $a^\dagger$. In this latter case $\lambda=1$,
	and the related vacuum $\tilde\varphi_0(x)$, solution of the equation $a^\dagger\tilde\varphi_0(x)=0$, is not square-integrable.

	\subsection{Back to non constant $m(x)$}\label{sectII2}

	It is interesting to check if there are other cases for which  $\varphi_0(x)$ in (\ref{210}) belongs to $\ltwo$, in particular for some $\lambda$ with positive (or zero) real part. Of course, in view of what we have just discussed, this is not possible for the harmonic oscillator. Something different is needed, and the role of a non-trivial dependence on $x$ in $m(x)$ is essential.
	
	We need to find conditions to have
	$$
	\|\varphi_0\|^2=\int_{\mathbb{R}}|\varphi_0(x)|^2dx=\abs{N_{\varphi,0}}^2\int_{\mathbb{R}}\sqrt{m(x)}\exp\left(\frac{\lambda_{R}}{\hbar^2}F^2(x)-2\gamma_RF(x)\right)\,dx<\infty,
	$$
	where $\lambda_{R}=\Re(\lambda)$ and $\gamma_{R}=\Re(\gamma)$. Since $m(x)$ is a positive function, and since $\dv{F(x)}{x}=\sqrt{m(x)}$, we deduce that $F(x)$ is a strictly increasing function. With the change of variable $x\rightarrow F(x)=: F$, we can rewrite
	\be
	\|\varphi_0\|^2=\abs{N_{\varphi,0}}^2\int_{F_{-\infty}}^{F_{+\infty}}\exp\left(\frac{\lambda_{R}}{\hbar^2}F^2-2\gamma_RF\right)\,dF,
	\label{210c}\en
	where we have introduced the notation $F_{\pm\infty}=F(\pm\infty)$. Their values depend on the explicit choice of $m(x)$. It may happen that they are both finite, both infinite or one finite and the other infinite. What is obvious is that, in any case, $F_{-\infty}<F_{+\infty}$. Different explicit situations will be considered in Section \ref{sect5}.  In particular, with the choice $m(x)=m_0e^{-x^2}$, $m_0$ a positive constant, we find that 	$F(x)=\sqrt{\frac{m_0\pi}{2}}\erf\left(\frac{x}{\sqrt{2}}\right)$, see (\ref{51}), so that $F_{-\infty}=-\sqrt{\frac{m_0\pi}{2}}$ and  $F_{\infty}=\sqrt{\frac{m_0\pi}{2}}$, see Section \ref{sectV1}. If we rather take, as in Section \ref{sectV2}, $m(x)=\frac{m_0}{1+x^2}$, $m_0>0$, then we get
	$F(x)=\sqrt{m_0}\arcsinh\left(x\right)$, see (\ref{58}). In this case we have $F_{-\infty}=-\infty$, while   $F_{\infty}=\infty$. In Section \ref{sectV3} we have also considered a third case, $m(x)=m_0\,e^x$, again with $m_0>0$. In this case $F(x)=2\sqrt{m_0}\,e^{\frac{x}{2}}$, so that $F_{-\infty}=0$, while   $F_{\infty}=\infty$.
	
	In the first case, when $F_{\infty}-F_{-\infty}<\infty$, the integral in (\ref{210c}) surely converges since it is the integral of a continuous function on a finite interval. Then $\varphi_0(x)\in\ltwo$ independently on the sign of $\lambda_R$, as well as of other details. 
	
	Let us then consider the second example in which $F_{\infty}-F_{-\infty}=\infty$. In this case we are (almost) forced to require that $\lambda_R<0$. Otherwise, the integral in (\ref{210c}) does not converge and $\varphi_0(x)\notin\ltwo$, at least if $F_{-\infty}=-\infty$ and   $F_{\infty}=\infty$. If one of these two quantities is finite, which is what happens for the third choice of $m(x)$ above, we can still have convergence of the integral in some interesting situation. In fact, let us consider the case of $\lambda_R=0$. In this case we have $\|\varphi_0\|^2=\abs{N_{\varphi,0}}^2\int_{F_{-\infty}}^{F_{+\infty}}e^{-2\gamma_RF}\,dF$, and it is clear that this integral exists finite in one of the following two cases: $\gamma_R>0$ and $F_{-\infty}$ finite, $-\infty<F_{-\infty}<F_{+\infty}\leq\infty$ (this is the case when $m(x)=m_0\,e^x$), or $\gamma_R<0$ and $F_{+\infty}$ finite,  $-\infty\leq F_{-\infty}<F_{+\infty}<\infty$. 
	
	The examples mentioned above for $m(x)$ suggest that different values for $F_{\pm\infty}$ are related to the asymptotic behavior of $m(x)$. In fact, it is easy to see that the only case in which  	$F_{\infty}-F_{-\infty}<\infty$, is when $m(x)$ goes to zero faster than $x^{-2}$ for $|x|\rightarrow\infty$.

	\vspace{2mm}

	After this general analysis, we want to compute the eigenvalue of $H$ corresponding to the eigenvector $\varphi_0(x)$ in (\ref{210}). A direct computation shows that
	\be
	H\varphi_0(x)=E_0\varphi_0(x), \qquad  E_0=-\frac{1}{2}\left(\gamma^2\hbar^2+\lambda\right),
	\label{211}\en
	which can be real or complex depending on the fact that the imaginary parts of $\gamma$ and $\lambda$  are zero or not. We notice that $\lambda$ enters also directly in the form of the eigenvalue of $H$.

	$H^\dagger$ is now the following:
	\begin{equation*}
		H^\dagger=-\frac{\hbar^2}{2m(x)}\dv[2]{}{x}+\frac{\hbar^2m'(x)}{2m^2(x)}\dv{}{x}+\frac{\hbar^2}{8}\left(\frac{m''(x)}{m^2(x)}-\frac{7(m'(x))^2}{4m^3(x)}\right)+\frac{\overline{\lambda}^2}{2\hbar^2}F^2(x)-\overline{\gamma\lambda}F(x).
	\end{equation*}
	It is clear that, if $\lambda=0$, then $H=H^\dagger$. But this also implies that $[H,A]=0$, see (\ref{22}), which, as already stressed, is not really what we want: $H$ and $A$ commute, and then $A$ is not really an ALO, in the usual sense, \cite{fern1,bagalo1}. Other (non trivial) cases in which, at least formally, $H=H^\dagger$, is when $\lambda\in\R\setminus\{0\}$ and $\gamma\in\R$, or when both $\lambda$ and $\gamma$ are purely imaginary.
	

	\subsection{Factorization of $H$}
	In \cite{bagalo1} it has been clarified that, if $H$ (or $H$ plus a constant) can be factorized, then the mathematical framework becomes richer and it is possibly easier to find its eigenstates and eigenvalues. This is essentially because whenever $H$ can be written as the product of two operators, then an algebraic condition on $H$ and $A$ which extends (\ref{22}), and which can be used in the construction of eigenvectors and eigenvalues of $H$, is always satisfied. However, we will not use this extended version of ALOs here. We refer to \cite{bagalo2} for further details in this direction.
	
	It is clear that, for self-adjoint Hamiltonians, $H_0=H_0^\dagger$, we should have a factorization of the form $H=D^\dagger D$, for some suitable operator $D$. If, on the other hand, $H\neq H^\dagger$, then the factorization could be of the form $H=GF$, with $G\neq F^\dagger$, in principle. This has also a side consequence, which can be interesting for its physical interpretation: going from $H_0$ to $H$ allows us to consider Hamiltonians which are not necessarily positive (or bounded from below).
	
	Our Hamiltonian $H$ in (\ref{21}) is not (even formally) self-adjoint, in general. For this reason we try to factorize $H$, or better $H-E_0\1$, as in $H-E_0\1=BA$, where $A$ is the one in (\ref{211}), while $B$ should still be computed (if it exists!).
	We assume that $B$ can be written as in (\ref{23}),
	\be
	B=\frac{1}{\sqrt{2}}\left(\eta(x)\dv{}{x}+\varepsilon(x)\right),
	\label{212}\en
	where $\eta(x)=\eta_R(x)+i\eta_I(x)$ and $\varepsilon(x)=\varepsilon_R(x)+i\varepsilon_I(x)$, as for $\alpha(x)$ and $\beta(x)$ in (\ref{23}). Adopting now the same strategy which produces the system in (\ref{24}), and with similar computations, we find here 
	\be
	B=\frac{\hbar^2}{\sqrt{2}}\left[-\frac{1}{ \sqrt{m(x)}}\dv{}{x}+ \frac{ m'(x)}{4 m^{3/2}(x)}  -\frac{\lambda}{\hbar^2}F(x)+\gamma\right].
	\label{213}\en
	It is clear that $B$ is densely defined and that, in particular, $B$ leaves $\scr$ stable, as $A$ does: if $f\in\scr$ then $Bf\in\scr$.
	
	\vspace{2mm}
	
	{\bf Remark:--} Notice that in general $B\neq A^\dagger$, since 
	\be
	\begin{split}
		A^\dagger &=\frac{1}{\sqrt{2}}\left[-\dv{}{x}\frac{1}{\sqrt{m(x)}}-\frac{ m'(x)}{4m^{3/2}(x)}-\frac{\overline{\lambda}}{\hbar^2}F(x)+\overline{\gamma}\right]=\\
		&=\frac{1}{\sqrt{2}}\left[-\frac{1}{\sqrt{m(x)}}\dv{}{x}+\frac{ m'(x)}{4m^{3/2}(x)}-\frac{\overline{\lambda}}{\hbar^2}F(x)+\overline{\gamma}\right].
	\end{split}
	\label{214}
	\en
	However,  if $\lambda\in\R\setminus\{0\}$ and $\gamma\in\R$, $B$ and $A^\dagger$ are indeed proportional: $B=\hbar^2A^\dagger$, and in particular they coincide if we work in units $\hbar=1$.

	The vacuum of $B^\dagger$, $B^\dagger\psi_0(x)=0$, can be found by solving the differential equation
	\begin{equation*}
		\frac{\hbar^2}{\sqrt{2}}\left[\frac{1}{\sqrt{m(x)}}\dv{}{x}-\frac{ m'(x)}{4m^{3/2}(x)}-\frac{\overline{\lambda}}{\hbar^2}F(x)+\overline{\gamma}\right]\psi_0(x)=0,
	\end{equation*}
	whose solution is
	\begin{equation}
		\psi_0(x)=N_{\psi,0}m^{\frac{1}{4}}(x)e^{\frac{\overline{\lambda}}{2\hbar^2}F^2(x)-\overline{\gamma} F(x)},
		\label{215}
	\end{equation}
	which is very similar to $\varphi_0(x)$ in (\ref{210}), with $\lambda$ and $\gamma$ replaced by their complex conjugates. The same analysis carried out for  $\varphi_0(x)$ can now be repeated for $\psi_0(x)$, to check if this is square integrable or not. The conclusions are indeed completely analogous to the previous ones, mostly related to the sign of $\lambda_R$, and will not be repeated here.

	The normalization constants $N_{\varphi,0}$ and $N_{\psi,0}$ could be (almost) fixed by requiring that $\langle\psi_0,\varphi_0\rangle=1$. We further require here that $N_{\varphi,0}=\overline{N_{\psi,0}}$. The general form of these normalization, and some comments, are given in Appendix C. 
	
	If we now compute $[A,B]$, we can check that this commutator is proportional to the identity operator $\1$. In particular we find that
	\be
	[A,B]=-\lambda\1,
	\label{216}\en
	where the commutator could be intended as acting, e.g., on functions of $\scr$. In terms of these operators we can write the Hamiltonian $H$  in \eqref{28}   as $H=BA+E_0\1$. Next, pseudo-bosonic operators, see \cite{bagspringerbook}, can be introduced by rescaling $A$ and $B$ as follows\footnote{Notice that the one we are adopting here is not the unique choice.}:
	\be
	a=-\frac{A}{\lambda}, \qquad b=B, \qquad \qquad [a,b]=\1, \qquad\mbox{ and }\qquad H=-\lambda ba+E_0\1=-\lambda N+E_0\1,
	\label{217}\en
	where $N=ba$ is the pseudo-bosonic number operator. Notice that $a$ is well defined since $\lambda\neq0$. Notice also that for  $\lambda\in\R\setminus\{0\}$ and $\gamma\in\R$, then $b^\dagger=-\lambda\hbar^2 a$. We see that in this case pseudo-bosons are not very different from ordinary bosons.
	The adjoint of $H$ is 
	\be
	H^\dagger=-\overline{\lambda}a^\dagger b^\dagger +\overline{E}_0\1=-\overline{\lambda}N^\dagger +\overline{E}_0\1.
	\label{218}\en
	
	Using the pseudo-bosonic approach, we can construct
	two families of (square-integrable) functions, $\F_\varphi=\{\varphi_n (x),n=0,1,2,...\}$ and $\F_\psi=\{\psi_n (x),n=0,1,2,...\}$,
	where
	\begin{equation}
		\label{219}
		\varphi_n(x)=\frac{b^n}{\sqrt{n!}}\varphi_0(x)=\frac{1}{\sqrt{n!}}\left( i\frac{\hbar\sqrt{\abs{\lambda}}}{\sqrt{2}}e^{\frac{i}{2}\theta_\lambda} \right)^{n}H_n\left(i\frac{\lambda F(x)-\hbar^2\gamma}{\hbar\sqrt{\abs{\lambda}}}e^{-\frac{i}{2}\theta_\lambda}\right)\varphi_0(x),
	\end{equation}
	and
	\begin{equation}
		\label{220}
		\psi_n(x)=\frac{(a^\dagger)^n}{\sqrt{n!}}\psi_0(x)=\frac{1}{\sqrt{n!}}\left( \frac{i}{\hbar\sqrt{2\abs{\lambda}}}e^{\frac{i}{2}\theta_\lambda } \right)^{n} H_n\left( i\frac{\hbar^2\overline{\gamma}-\overline{\lambda} F(x) }{\hbar\sqrt{\abs{\lambda}}}e^{\frac{i}{2} \theta_\lambda } \right)\psi_0(x).
	\end{equation}
	We refer to Appendix B for the details of this derivation. Here we have used the polar form for $\lambda$: $\lambda=|\lambda|e^{i\theta_\lambda}$, $\theta_\lambda\in[0,2\pi[$. Hence $\lambda>0$ if $\theta_\lambda=0$, while  $\lambda<0$ if $\theta_\lambda=\pi$. Also, since $\lambda=|\lambda|\left(\cos\theta_\lambda+i\,\sin\theta_\lambda\right)$, $\lambda_R<0$ if $\theta_\lambda\in]\frac{\pi}{2},\frac{3\pi}{2}[$.

	It is clear that, if $\varphi_0(x)$ and $\psi_0(x)$ are in $\ltwo$ (see our previous analysis) then all the other functions in (\ref{219}) and (\ref{220}) are square-integrable as well. This is clear if $F_{\infty}-F_{-\infty}<\infty$. If we rather have $F_{\infty}-F_{-\infty}=\infty$, for instance if $F_{\infty}=-F_{-\infty}=\infty$, our claim follows from the fact that $|H_n|^2$ diverges in its argument polynomially, while $|\varphi_0(x)|^2$ goes to zero exponentially. The other cases can be analysed in a similar way.
	

	The following ladder and eigenvalue equations have been deduced
	for the elements of $\F_\varphi$ and $\F_\psi$:
	
	\begin{align}
		\label{221}
		\begin{cases}
			b\varphi_n=\sqrt{n+1}\varphi_{n+1}   &\quad n \geq 0, \\
			a\varphi_0=0, \qquad a\varphi_n=\sqrt{n}\varphi_{n-1}, &\quad n \geq 1, \\
			a^{\dagger}\psi_n=\sqrt{n+1}\psi_{n+1}, &\quad  n \geq 0, \\
			b^{\dagger}\psi_0=0, \qquad b^{\dagger}\psi_n=\sqrt{n}\psi_{n-1}, &\quad  n \geq 1,
		\end{cases}
	\end{align}
	as well as
	\begin{equation}
		\label{222}
		N\varphi_k(x)=k\varphi_k(x),\qquad N^\dagger\psi_k(x)=k\psi_k(x),
	\end{equation}
	for all $k=0,1,2,\ldots.$ Then, due to expressions (\ref{217}) and (\ref{218}) for $H$ and $H^\dagger$, it follows  that the $\varphi_n(x)$ are eigenstates of $H$ with eigenvalues $E_n$,
	\begin{equation}
		\label{223}
		H\varphi_n(x)=E_n\varphi_n(x),
	\end{equation}
	where 
	\begin{equation}
		\label{224}E_n=E_0-n\lambda=-\frac{\gamma^2\hbar^2}{2}-\left(n+\frac{1}{2}\right)\lambda,
	\end{equation}
	and that the $\psi_n(x)$ are eigenstates of $H^\dagger$ with eigenvalues $\overline{E_n}$:
	\begin{equation}
		\label{225}
		H^\dagger\psi_n(x)=\overline{E_n}\psi_n(x).
	\end{equation}
	As expected, the eigenvalues of $H$ and $H^\dagger$ could be real or complex, depending on the nature of $\lambda$. The role of $\gamma$ is not as crucial, since $\gamma$ is simply an integration constant, while $\lambda$ is one of the key ingredients in our framework, see (\ref{22}). In particular, $H$ and $H^\dagger$ have the same eigenvalues if $\lambda,\gamma\in\mathbb{R}$. Otherwise they do not.
	
	\vspace{2mm}
	
	{\bf Remark:--} In order to understand our results, it may be useful to summarize what we have done so far: with the aim of finding the eigensystem of $H$ in (\ref{21}) we have adopted first the strategy connected with ALOs, trying to solve (\ref{22}), i.e., to identify the expressions of $V(x)$, $A$ and $\lambda$ which are compatible with (\ref{22}). Then we have imposed a second constraint on $H$, i.e. to be (almost) factorizable in the following sense: $H-E_0\1=BA$. Not surprisingly, these requirements restrict the forms of $V(x)$ which fit all our requirements, and in fact what we have found is that $H$ and $H^\dagger$ automatically are connected with pseudo-bosons. For this reason it is not a big surprise that $E_n$ is linear in $n$, see (\ref{224}). To get something different we should relax some of our assumptions, and look for a (possibly larger) class of solvable Hamiltonians. This is part of our plans for future work.
	
	\vspace{2mm}

	The functions in 		$\F_\varphi$ and $\F_\psi$, under the conditions that they are all in $\ltwo$, are biorthonormal. This is a standard consequence of (\ref{222}). In the literature it is discussed that biorthonormal sets need not to be bases for $\Hil$, or even total. This, in fact, can also be a non trivial aspect for orthonormal sets. What we will do now is to check that $\F_\varphi$ and $\F_\psi$ are indeed rather {\em rich} sets. In particular, we show	
	first that $\F_{\varphi}$ is total in $\ltwo$, i.e. that the only square-integrable function which is orthogonal to all the $\varphi_n(x)$ is the zero function. 
	
	First we observe that $\F_{\varphi}$ is total if and only if the set
	$$
	\F_\eta=\left\{\eta_n(x)=(F(x))^n(m(x))^{1/4}\exp\left(\frac{\lambda}{2\hbar^2}F^2(x)-\gamma F(x)\right)\right\}
	$$
	is total in $\ltwo$. The proof of this claim is not particularly different from what is done in similar contexts, see \cite{bag2010}, and later, in \cite{baginbagbook}, and will not be repeated here: we are essentially replacing Hermite polynomials with monomials. 
	
	Le us now consider a function $f(x)\in\ltwo$ which is orthogonal to all the $\eta_n(x)$: $\pin{f}{\eta_n}=0$, for all $n=0,1,2,\ldots$. With the same change of variable which produces (\ref{210c}) we get
	$$
	\int_{F_{-\infty}}^{F_{+\infty}}\overline{g(x(F)})\,F^n\,\exp\left(\frac{\lambda}{2\hbar^2}F^2-\gamma\,F\right)\,dF=0,
	$$
	$\forall n\geq0$. Here $g(x)=\frac{f(x)}{(m(x))^{1/4}}$. We recall that $F(x)$ is monotone, and therefore it is invertible.
	
	We use now a general result given in \cite{kolfom}: {\em let $h(x)$ be a measurable function on $(a,b)$, $-\infty\leq a<b\leq +\infty$, $h(x)\neq0$ almost everywhere in $(a,b)$. Let us assume that $|h(x)|\leq C\,e^{-\delta |x|}$, for some $C, \delta>0$, then the set $\{x^n h(x), \,n=0,1,2,\ldots\}$ is total in $\ltwo$.}
	
	It is clear that the function $\exp\left(\frac{\lambda}{2\hbar^2}F^2-\gamma\,F\right)$, if $\lambda_R<0$, satisfies the above requirement for $h(x)$. Then we conclude that $g(x(F))=0$ a.e. in $F$, which implies that $g(x)=0$ a.e. in $\mathbb{R}$. Hence $\F_\varphi$ is total in $\ltwo$. A similar proof can be repeated for $\F_\psi$, which is also total in $\ltwo$.
	
	What remains open is whether these two sets are bases or not or, which is usually not so easy to understand. On the other hand, it is not hard to check that they are $\G$-quasi bases, \cite{bagspringerbook,baginbagbook}, where $\G=l.s.\{\varphi_n(x)\}\cap l.s.\{\psi_n(x)\}$, at least if $\G$ is dense in $\ltwo$. Here "l.s." stands for {\em linear span}. For instance, $l.s.\{\varphi_n(x)\}$ is the set of all the finite linear combinations of the functions $\varphi_n(x)$. In fact, in this case we immediately get 
	\be
	\sum_n\pin{f}{\varphi_n}\pin{\psi_n}{g}=\sum_n\pin{f}{\psi_n}\pin{\varphi_n}{g}=\pin{f}{g},
	\label{add1}\en
	$\forall f,g\in\G$. In most situations considered in the literature the two sets are not bases, so our guess is that $\F_\varphi$ and $\F_\psi$ are quasi bases, but not bases.

	\section{Bi-coherent states}\label{sect3}
	We will now  analyze bi-coherent states associated to our lowering operators $a$ and $b^\dagger$. In particular, in what follows we will consider two alternative approaches.  The first relies on the fact that bi-coherent states, here denoted by $\varphi(z;x)$ and $\psi(z;x)$, are eigenstates of lowering operators $a$ and $b^\dagger$ with complex eigenvalue $z$. In the second case, bi-coherent states ($\tilde{\varphi}(z;x)$ and $\tilde{\psi}(z;x)$) are defined as the sum of certain, always convergent, power series in $z$. Finally, we will compare them to see if they are equivalent, i.e. if the two procedures produce the same states. This analysis follows the same main steps considered in \cite{bagspringerbook} in dealing with bi-coherent states. We will also briefly comment on the use of some deformed displacement operators.
	
	\subsection{First approach}
	We first solve the differential equations in	
	\begin{equation}
		\label{31}
		a\varphi(z;x)=z\varphi(z;x),\qquad \text{and}\qquad b^\dagger\psi(z;x)=z\psi(z;x),
	\end{equation}
	for all $z\in\mathbb{C}$, using the differential expression for $a$ and $b^\dagger$. Simple calculations produce
	\begin{align}
		\varphi(z;x)&=\mathcal{N}_{\varphi,z}\sqrt[4]{m(x)}e^{\frac{\lambda}{2\hbar^2} F^2(x)-(z\lambda\sqrt{2}+\gamma)F(x)}=\mathcal{M}_{\varphi,z}e^{-z\lambda\sqrt{2}F(x)}\varphi_0(x),\label{32}\\
		\psi(z;x)&=\mathcal{N}_{\psi,z}\sqrt[4]{m(x)}e^{\frac{\overline{\lambda}}{2\hbar^2} F^2(x)+\left(\frac{z\sqrt{2}}{\hbar^2}-\overline{\gamma}\right)F(x)}=\mathcal{M}_{\psi,z}e^{\frac{z\sqrt{2}}{\hbar^2}F(x)}\psi_0(x),\label{33}
	\end{align}
	where $\mathcal{N}_{\varphi,z}$ and $\mathcal{N}_{\psi,z}$  are normalization  constants that depend on  $z$, $\mathcal{M}_{\varphi,z}=\frac{\mathcal{N}_{\varphi,z}}{N_{\varphi,0}}$ and $\mathcal{M}_{\psi,z}=\frac{\mathcal{N}_{\psi,z}}{N_{\psi,0}}$. We see that there exists a strong relation between, say  $\varphi(z;x)$ and $\varphi_0(x)$, and this is independent on the specific choice of $m(x)$, in particular.

	\subsection{Second approach}
	
	We consider here the following $z$-dependent series:
	\begin{equation}
		\label{37}
		\tilde{\varphi}(z;x)=e^{-\frac{\abs{z}^2}{2}}\sum_{n=0}^\infty\frac{z^n}{\sqrt{n!}}\varphi_n(x),\qquad \tilde{\psi}(z;x)=e^{-\frac{\abs{z}^2}{2}}\sum_{n=0}^\infty\frac{z^n}{\sqrt{n!}}\psi_n(x).
	\end{equation}
	
	In \cite{bagspringerbook} it has been proposed a detailed way to check that these series are indeed convergent, in some disk of the complex plane or in the whole plane. This method is based on the estimates of $\|\varphi_n\|$ and $\|\psi_n\|$ and it has been proved that, for pseudo-bosons under very mild assumptions\footnote{This claim is not true for weak pseudo-bosons. But this is not the case here.} on $\|\varphi_n\|$ and $\|\psi_n\|$, these estimates guarantee that both series in (\ref{37}) converge in all of $\mathbb{C}$. It has also been proven that the eigenvalue equations in (\ref{31}) are both satisfied by the vectors in (\ref{37}), so that it is natural to ask if the functions in (\ref{32}) and (\ref{33}) coincide with the sum of the series in (\ref{37}). Indeed, as we will show shortly, this is the case.

	Before checking this claim, we briefly recall what is behind the series in (\ref{37}). For that we introduce the following formal displacement operators 
	
	
	\begin{equation}
		\label{34}
		V(z)=e^{zb-\overline{z}a}=e^{-\frac{\abs{z}^2}{2}}e^{zb}e^{-\overline{z}a},
	\end{equation}
	
	and

	\begin{equation}
		\label{35}
		W(z)=e^{za^\dagger-\overline{z}b^\dagger}=e^{-\frac{\abs{z}^2}{2}}e^{za^\dagger}e^{-\overline{z}b^\dagger}.
	\end{equation}
	These are {\em formal} in the sense that they are both unbounded, and differ from the {\em standard} displacement operator since they are not unitary. Also, in (\ref{34}) and (\ref{35}) the Baker-Campbell-Hausdorff formula has been used, but this is far from being a trivial task, if we want to use mathematical rigor. We refer to \cite{bagspringerbook} for a detailed analysis of this specific problem. In this few lines, we just use formal computations since the role of the operators  $V(z)$ and $W(z)$, although being interesting, is really minimal for us, here. They both admit inverse and it can be shown that
	
	\begin{equation}
		\label{36}
		V^\dagger(z)=W^{-1}(z),\qquad W^\dagger(z)=V^{-1}(z),\qquad \forall z\in\mathbb{C}.
	\end{equation}

	Then, because of \eqref{34} and \eqref{35}, we have 
	\begin{equation}
		\label{37b}
		\begin{split}
			V(z)\varphi_0(x)&=e^{-\frac{\abs{z}^2}{2}}e^{zb}e^{-\overline{z}a}\varphi_0(x)=e^{-\frac{\abs{z}^2}{2}}e^{zb}\varphi_0(x)=\\
			=&e^{-\frac{\abs{z}^2}{2}}\sum_{n=0}^\infty\frac{z^n}{n!}b^n\varphi_0(x)=e^{-\frac{\abs{z}^2}{2}}\sum_{n=0}^\infty\frac{z^n}{\sqrt{n!}}\varphi_n(x)=	\tilde{\varphi}(z;x),
		\end{split}
	\end{equation}
	and
	\begin{equation}
		\label{38b}
		\begin{split}
			W(z)\psi_0(x)&=e^{-\frac{\abs{z}^2}{2}}e^{za^\dagger}e^{-\overline{z}b^\dagger}\psi_0(x)=e^{-\frac{\abs{z}^2}{2}}e^{za^\dagger}\psi_0(x)=\\
			=&e^{-\frac{\abs{z}^2}{2}}\sum_{n=0}^\infty\frac{z^n}{n!}(a^\dagger)^n\psi_0(x)=e^{-\frac{\abs{z}^2}{2}}\sum_{n=0}^\infty\frac{z^n}{\sqrt{n!}}\psi_n(x)=\tilde{\psi}(z;x).
		\end{split}
	\end{equation}
	In other words, we can look at the series in (\ref{37}) as the result of the action of $V(z)$ on $\varphi_0(x)$, and of $W(z)$ on $\psi_0(x)$.

	\subsection{Comparison of approaches}
	
	We check here if the states  in \eqref{37}  are equivalent to those  previously defined in \eqref{32} and \eqref{33}, $\tilde{\varphi}(z;x)=\varphi(z;x)$ and $\tilde{\psi}(z;x)=\psi(z;x)$.
	
	We replace in \eqref{37} the expression of $\varphi_n(x)$ in \eqref{221}

	\begin{equation}
		\begin{split}
			\label{39}
			\tilde{\varphi}(z;x)&=e^{-\frac{\abs{z}^2}{2}}\sum_{n=0}^\infty\frac{z^n}{\sqrt{n!}}\varphi_n(x)=\\
			&=e^{-\frac{\abs{z}^2}{2}}\varphi_0(x)\sum_{n=0}^\infty\frac{z^n}{n!}\left(i\frac{\hbar\sqrt{\abs{\lambda}}}{\sqrt{2}}e^{\frac{i}{2}\theta_\lambda }\right)^{n}H_n\left(i\frac{\lambda F(x)-\hbar^2\gamma}{\hbar\sqrt{\abs{\lambda}}}e^{-\frac{i}{2}\theta_\lambda}\right).
		\end{split}
	\end{equation}
	Looking at the previous equation and using  the exponential generating function of the Hermite polynomials	
	\begin{equation*}
		e^{2\xi t-t^2}=\sum_{n=0}^\infty H_n(\xi)\frac{t^n}{n!},
	\end{equation*}
	we notice that in this case the parameter $t$ and the variable $\xi$ are
	\begin{equation*}
		t=iz \frac{\hbar\sqrt{\abs{\lambda}}}{\sqrt{2}}e^{\frac{i}{2}\theta_\lambda }\qquad\text{and}\qquad \xi=i\frac{\lambda F(x)-\hbar^2\gamma}{\hbar\sqrt{\abs{\lambda}}}e^{-\frac{i}{2}\theta_\lambda},
	\end{equation*}
	the  products $t^2$ and $2\xi t$ are
	\begin{equation*}
		t^2=-z^2 \frac{\hbar^2\abs{\lambda}}{2}e^{i\theta_\lambda }=-\frac{z^2\hbar^2\lambda}{2} \qquad\text{and}\qquad 2\xi t=2z\frac{\hbar^2\gamma-\lambda F(x)}{\sqrt{2}}=z\sqrt{2}\left(\hbar^2\gamma-\lambda F(x)\right).
	\end{equation*}
	By replacing them, we have
	\begin{equation}
		\label{310}
		\begin{split}
			\tilde{\varphi}(z;x)&=e^{-\frac{\abs{z}^2}{2}}e^{z\sqrt{2}\left(\hbar^2\gamma-\lambda F(x)\right)+\frac{z^2\hbar^2\lambda}{2}}\varphi_0(x)=\\&=e^{-\frac{\abs{z}^2}{2}+z\sqrt{2}\hbar^2\gamma+\frac{z^2\hbar^2\lambda}{2}}e^{-z\sqrt{2}\lambda F(x)}\varphi_0(x)=\varphi(z;x)
		\end{split}
	\end{equation}
	with \be \mathcal{M}_{\varphi,z}=\exp{-\frac{\abs{z}^2}{2}+z\sqrt{2}\hbar^2\gamma+\frac{z^2\hbar^2\lambda}{2}}.\label{310b}\en Therefore, $\tilde{\varphi}(z;x) \equiv \varphi(z;x)$, with the normalization constant $\mathcal{M}_{\varphi,z}$ now explicitly determined. Notice that the convergence holds in all the complex plane. A similar calculation for $\tilde{\psi}(z;x)$ leads to the same conclusion of equivalence with $\psi(z;x)$. In particular, through a simple check, we can see that 
	\begin{equation}\label{310c}
		\mathcal{M}_{\psi,z}=\exp{-\frac{\abs{z}^2}{2}-z\sqrt{2}\overline{\left(\frac{\gamma}{\lambda}\right)}+\frac{z^2}{2\hbar^2\overline{\lambda}}}.
	\end{equation}
	With standard technique, see \cite{bagspringerbook} for instance, we conclude further that
	$$
	\frac{1}{\pi}\int_\mathbb{C}\langle f,\varphi(z;x)\rangle\langle \psi(z;x),g\rangle dz=\langle f,g\rangle, 
	$$
	for all  $ f,g\in\G$, which is the resolution of the identity for our states. Here $\G$ is the set we have defined previously, in connection with \eqref{add1}.

	{
		
		\vspace{2mm}
		
		{\bf Remark:--} It is interesting, also in view of the plots we will show later, to see what happens when we work under the conditions $\lambda\in\R\setminus\{0\}$ and $\gamma\in\R$. In this case, we have already commented that $b^\dagger=-\lambda\hbar^2 a$. For this reason we expect that the bi-coherent states in (\ref{31}) essentially coincide. In fact, this is not exactly so. However, it is easy to check that there is a strong relation between $\varphi(z;x)$ and $\psi(z;x)$. Indeed, since $a\varphi(z;x)=z\varphi(z;x)$, and since $b^\dagger=-\lambda\hbar^2 a$, it follows that
		\be
		\psi(z;x)=\varphi\left(-\frac{z}{\hbar^2\lambda},x\right).
		\label{310d}
		\en
		In fact a simple computation shows that, with this choice, $b^\dagger\psi(z;x)=z\psi(z;x)$. However, the choice in (\ref{310d}) differs from what we get from (\ref{32}) and (\ref{33}) if we further fix the normalizations as in (\ref{310b}) and (\ref{310c}). For this reason we replace (\ref{310d}) with
		\be
		\psi(z;x)=\alpha(z,\lambda)\varphi\left(-\frac{z}{\hbar^2\lambda},x\right),
		\label{310e}
		\en
		where
		\begin{equation}
			\alpha(z,\lambda)=\exp\left\{\frac{\abs{z}^2}{2}\left(\frac{1}{\abs{\lambda}^2\hbar^4}-1\right)\right\}.
		\end{equation}
		Of course also the state in (\ref{310e}) satisfies the eigenvalue equation $b^\dagger\psi(z;x)=z\psi(z;x)$.

	}

		\section{Alternative Approach}\label{sect4}

		In this section we propose an alternative approach to the analysis of the Hamiltonians in (\ref{21}) and in (\ref{28}). 
		
		First of all we introduce a new (complex) variable $y=F(x)-\frac{\hbar^2\gamma}{\lambda}$. In fact, since $F(x)$ is real, $y$ is real or not depending on the imaginary part of $\frac{\hbar^2\gamma}{\lambda}$. Consider the shifted real line 
		\begin{equation}
			\mathbb{R}_s=\mathbb{R}-i \Im{\frac{\hbar^2 \gamma}{\lambda}}=\left\{ x-i \Im{\frac{\hbar^2 \gamma}{\lambda}}, \quad x \in \mathbb{R}\right\}
		\end{equation}
		and the space $C^{\infty}(\mathbb{R}_s)$ of functions $f:\mathbb{R}_s \rightarrow \mathbb{C}$, infinitely differentiable in $x$. The functions of this space can be simply written as $f\left(x-\frac{\hbar^2\gamma}{\lambda}\right)$ where $x\in \mathbb{R}$.

		The Hamiltonian (\ref{28}) is a  well-defined operator acting on $C^{\infty}(\mathbb{R}_s)$. We want to write this Hamiltonian in terms of $y$ variables. Recalling that $F(x)$ is always increasing, and calling $R(y)=\left(F_{-\infty}-\frac{\hbar^2 \gamma}{\lambda},F_{\infty}-\frac{\hbar^2 \gamma}{\lambda}\right)$ the range of $y(x)$, it is clear that the inverse map of $y(x)$ exists. We call this inverse map   $y^{-1}(x)$ and we observe that
		\begin{equation}
			y^{-1}(x):\left(F_{-\infty}-\frac{\hbar^2 \gamma}{\lambda},F_{\infty}-\frac{\hbar^2 \gamma}{\lambda}\right)\longrightarrow \mathbb{R}.
		\end{equation}
		If we now rewrite $H$ in (\ref{28}) in terms of $y$ we find that $H$ assumes the following expression: 
		\be
		H=\frac{\hbar^2}{2}\left[-\dv[2]{}{y}+\frac{m'(y)}{2m^2(y)}\dv{}{y}+\frac{1}{4}\left(\frac{m''(y)}{m(y)}-\frac{5(m'(y))^2}{4m^2(y)}\right)+\frac{\lambda^2}{\hbar^4}y^2-\gamma^2\right].
		\label{41}\en
		Incidentally we observe that, since \(y\) is real up to a complex shift,  the momentum operator $p_y=-i\hbar\dv{}{y}$ remains Hermitian, whereas the position operator $y$ is not. In fact, $y$ is Hermitian up to a shift (which could be real or not).
		
		The Hamiltonian (\ref{41}) can be connected to a quadratic Hamiltonian with the following transformation: rewriting the (generic) eigenstate of $H$, $\tilde{\varphi}(y)$, as $\tilde{\varphi}(y)=e^{\chi(y)}\varphi(y)$, and assuming further that
		
		\begin{equation}
			\chi(y)=\frac{1}{4}\ln(m(y)),
		\end{equation}
		it follows that
		\begin{equation}\label{45}
			H\tilde{\varphi}(y)=m^{\frac{1}{4}}(y)\left(-\frac{\hbar^2}{2}\dv[2]{}{y}+\frac{e^{2i\theta}}{2}\left(\frac{\abs{\lambda}}{\hbar}\right)^2 y^2-\frac{\hbar^2\gamma^2}{2}\right)\varphi(y),
		\end{equation}
		which relates in an easy way $H$ with a  quadratic Hamiltonians times $m^{\frac{1}{4}}(y)$. Here, to slightly simplify the notation, we have used $\theta=\theta_\lambda-\pi$, so that  $\lambda=-\abs{\lambda}e^{i\theta}$. 
		Let us now define $\Omega=\frac{\abs{\lambda}}{\hbar}$. Hence
		\begin{equation}
			\label{48}
			H\tilde{\varphi}(y)=m^{\frac{1}{4}}(y)\left[-\frac{\hbar^2}{2}\dv[2]{}{y}+\frac{e^{2i\theta}\Omega^2}{2} y^2-\frac{\hbar^2\gamma^2}{2}\right]\varphi(y)=m^{\frac{1}{4}}(y)\left(H_{\theta}-\frac{\hbar^2\gamma^2}{2}\right)\varphi(y)
		\end{equation}
		where $H_{\theta}$ is
		\be
		H_\theta=\frac{1}{2}\left(p_y^2+e^{2i\theta}\Omega^2y^2\right),
		\label{48b}\en
		the Hamiltonian analyzed in many details in \cite{bagiqho}, in connection with the inverted harmonic oscillator. It is not hard to prove, see also \cite{bagbaku}, that  calling $H_{HO}$, $H_{IHO}$ and $H_{BK}$ respectively the Hamiltonians of harmonic oscillator, of the inverted harmonic oscillator, and the so-called Berry-Keating Hamiltonian, 
		$$
		H_{IHO}=\frac{1}{2}\left(p_y^2-\Omega^2\hat y^2\right), \quad H_{HO}=\frac{1}{2}\left(p_y^2+\Omega^2 y^2\right), \quad H_{BK}=\frac{1}{2}(p_y\,y+yp_y), 
		$$
		the following equality holds (in the sense of unbounded operators):
		\begin{equation}
			e^{-\frac{\theta}{2\hbar}H_{BK}}H_{HO}e^{\frac{\theta}{2\hbar}H_{BK}}=H_{HO}\cos(\theta)-iH_{IHO}\sin({\theta}).
		\end{equation}
		Using further the identity
		\begin{equation}
			H_{HO}\cos(\theta)-iH_{IHO}\sin({\theta})=e^{-i\theta}H_{\theta},
		\end{equation}
		we deduce that
		\begin{equation}
			H_{\theta}=e^{i\theta}e^{-\frac{\theta}{2\hbar}H_{BK}}H_{HO}e^{\frac{\theta}{2\hbar}H_{BK}}.
		\end{equation}
		
		%
		%
		Equation \eqref{48} provides also information about the eigenstates and eigenvalues of \(H\). If $H_{\theta}\varphi_n^{(\theta)}=E_n^{(\theta)}\varphi_n^{(\theta)}$, it follows that $\tilde{\varphi}_n(y)=N^{\varphi}_{n}m^{\frac{1}{4}}(y)\varphi_n^{(\theta)}(y)$ is an eigenstate of $H$ with eigenvalue $E_n^{(\theta)}-\frac{\hbar^2\gamma^2}{2}$.
		As shown in \cite{bagiqho} the eigenstates of $H_{\theta}$ are:
		\begin{equation}
			\varphi_n^{(\theta)}(y)=\frac{N^{(\theta)}}{\sqrt{2^nn!}}H_{n}\left(e^{i\theta/2}\sqrt{\frac{\Omega}{\hbar}}y\right)e^{-\frac{\Omega}{2\hbar}e^{i\theta}y^2}
			\label{eigenstate_swansom_like}
		\end{equation}
		where $N^{(\theta)}=\left(\frac{\Omega}{\hbar\pi}\right)^{1/4}e^{i\theta/4}$ with corresponding eigenvalues $E_n^{(\theta)}=\Omega e^{i \theta}\left(n+\frac{1}{2}\right)$.
		This implies that the eigenstates of \(H\) are
		\begin{equation}
			\tilde{\varphi}_n(y)=\frac{N^{\varphi}_{n}N^{(\theta)}}{\sqrt{2^nn!}}H_{n}\left(e^{i\theta/2}\sqrt{\frac{\Omega}{\hbar}}y\right)m^{\frac{1}{4}}(y)e^{-\frac{\Omega}{2\hbar}e^{i\theta}y^2}
		\end{equation}
		with eigenvalues 
		\begin{equation}
			\begin{split}
				E_n=&E_n^{(\theta)}-\frac{\hbar^2\gamma^2}{2}=\hbar\Omega e^{i \theta}\left(n+\frac{1}{2}\right)-\frac{\hbar^2\gamma^2}{2}=\abs{\lambda} e^{i \theta}\left(n+\frac{1}{2}\right)-\frac{\hbar^2\gamma^2}{2}=\\=&-\lambda\left(n+\frac{1}{2}\right)-\frac{\hbar^2\gamma^2}{2}.
			\end{split}
		\end{equation}
		We see that eigenvalues are the same as those derived earlier. Also, if we put $N^\varphi_n=\left( i\frac{\hbar\sqrt{\abs{\lambda}}}{\sqrt{2}}e^{\frac{i}{2}\theta_\lambda} \right)^{n}$, returning to the original variable $x$, we obtain the eigenstates in \eqref{219}. A similar procedure can be applied for the states $\psi_n(x)$,  paying attention to the fact that the argument of the wave functions are, in general, complex. Hence the eigenstates of $H^{\dagger}$ are proportional to $m(\overline{y})\psi^{(\theta)}_{n}(\overline{y})$, where we recall that $y=F(x)-\frac{\hbar^2\gamma}{\lambda}$.

			\section{Different choices of the $m(x)$}\label{sect5}
			In this section, different choices for the function $m(x)$ in $C^{\infty}(\mathbb{R})$ are considered, and the square-integrability of the corresponding eigenfunctions is analyzed. Also, some plots for bi-coherent states are given, showing the differences between $\varphi(z;x)$ and $\psi(z;x)$.  
			
			\subsection{case 1: $m(x)=m_0e^{-x^2}$}\label{sectV1}
			In this case, the mass function $m(x)$ decreases very rapidly. For this reason, according to our previous analysis, we expect the eigenfunctions to be square-integrable regardless of the sign of $\lambda_R$. The function $F(x)$ is
			\begin{equation}
				\label{51}
				F(x)=\sqrt{m_0}\int_0^x e^{-t^2/2}dt=\sqrt{\frac{m_0\pi}{2}}\erf\left(\frac{x}{\sqrt{2}}\right)
			\end{equation}
			where $\erf(x)$ is the error function. As already remarked, this function it is monotonically increasing and bounded from below and from above.
			{With this choice of mass, the potential $V(x)$ in (\ref{21}), as given by \eqref{27}, is
				\begin{equation}
					V(x)=-\frac{\hbar^2e^{x^2}}{4m_0}\left(\frac{3}{2}x^2+1\right)+\frac{\lambda^2 \pi m_0}{4 \hbar^2}\erf^2\left(\frac{x}{\sqrt{2}}\right)-\lambda \gamma\sqrt{\frac{m_0 \pi}{2}}\erf\left(\frac{x}{\sqrt{2}} \right).
				\end{equation}
			}
			
			The Hamiltonian and the two pseudo-bosonic operators are given by
			{
				\begin{equation}
					\label{52}
					H=-\frac{\hbar^2e^{x^2}}{2m_0}\dv{^2}{x^2}-\frac{\hbar^2xe^{x^2}}{m_0}\dv{}{x}-\frac{\hbar^2e^{x^2}}{4m_0}\left(\frac{3}{2}x^2+1\right)+\frac{\lambda^2 \pi m_0}{4 \hbar^2}\erf^2\left(\frac{x}{\sqrt{2}}\right)-\lambda \gamma\sqrt{\frac{m_0 \pi}{2}}\erf\left(\frac{x}{\sqrt{2}}\right),
			\end{equation}}
			\begin{align}
				a &=-\frac{1}{\lambda \sqrt{2m_0}}\left[e^{x^2/2}\left(\dv{}{x}+\frac{x}{2}\right)-\frac{\lambda m_0}{\hbar^2}\sqrt{\frac{\pi}{2}}\erf\left(\frac{x}{\sqrt{2}}\right)  +\gamma \sqrt{m_0}\right], \label{53}  \\
				b &=-\frac{\hbar^2}{\sqrt{2m_0}}\left[e^{x^2/2}\left(\dv{}{x}+\frac{x}{2}\right)+\frac{\lambda m_0}{\hbar^2}\sqrt{\frac{\pi}{2}}\erf\left(\frac{x}{\sqrt{2}}\right)  -\gamma \sqrt{m_0}\right].\label{54}
			\end{align}
			The ground eigenfunctions are
			\begin{align}
				\varphi_0(x)&=N_{\varphi,0}m_0^{\frac{1}{4}}\exp\left(-\frac{x^2}{4}+\frac{\lambda \pi m_0}{4\hbar^2}\erf^2\left(\frac{x}{\sqrt{2}}\right)-\gamma\sqrt{\frac{m_0 \pi}{2}}\erf\left(\frac{x}{\sqrt{2}}\right)\right),\label{55}\\
				\psi_0(x)&=N_{\psi,0}m_0^{\frac{1}{4}}\exp\left(-\frac{x^2}{4}+\frac{\overline{\lambda} \pi m_0}{4\hbar^2}\erf^2\left(\frac{x}{\sqrt{2}}\right)-\overline{\gamma}\sqrt{\frac{m_0 \pi}{2}}\erf\left(\frac{x}{\sqrt{2}}\right)\right).\label{56}
			\end{align}
			{
				By using (\ref{norm_constant_general_case}) and the fact that the $\erf(z)$ is an odd function, the normalization constant is   
			
			\begin{equation}
				\begin{split}
					N_{\varphi,0}=\overline{N_{\psi,0}}=&\left[\frac{i\hbar}{2}\sqrt{\frac{\pi}{\abs{\lambda}}}e^{-\frac{\hbar^2\gamma^2}{\lambda}-i\frac{\theta_{\lambda}}{2}} \left[\erf\left(i\frac{e^{i\frac{\theta_{\lambda}}{2}}\sqrt{\abs{\lambda}}}{\hbar}\left\{\sqrt{\frac{m_0\pi}{2}}-\frac{ \hbar^2\gamma}{\lambda}\right\}\right)+\right.\right.\\
					&+\left.\left.\erf\left(i\frac{e^{i\frac{\theta_{\lambda}}{2}}\sqrt{\abs{\lambda}}}{\hbar}\left\{\sqrt{\frac{m_0\pi}{2}}+\frac{ \hbar^2\gamma}{\lambda}\right\}\right) \right]\right]^{-\frac{1}{2}}.
				\end{split}
		\end{equation}}
		
		When $x\rightarrow\pm \infty$, the functions above behave like $e^{-x^2/4}$, and are square integrable,  as expected. 
		This result suggests that a rapidly decaying mass acts as a strong confining potential, ensuring that the states belong to $\ltwo$ regardless of the sign of $\lambda_R$. This is in agreement with what we have deduced in our analysis in Section \ref{sect2}.
		
\begin{figure}[ht!]
			\centering
			\begin{subfigure}[b]{0.45\textwidth}
				\centering
				\includegraphics[width=\textwidth]{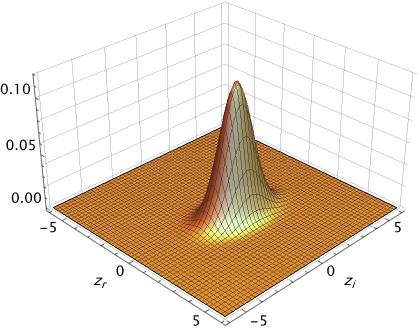}
			\end{subfigure}
			\hfill
			\begin{subfigure}[b]{0.45\textwidth}
				\centering
				\includegraphics[width=\textwidth]{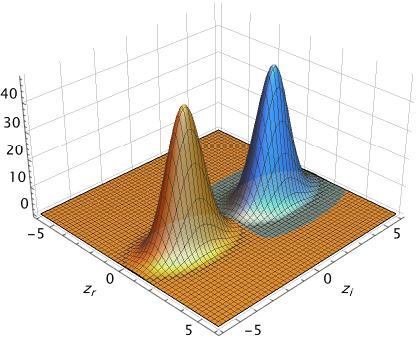}
			\end{subfigure}
			
			\vspace{0.5cm} 
			
			\begin{subfigure}[b]{0.45\textwidth}
				\centering
				\includegraphics[width=\textwidth]{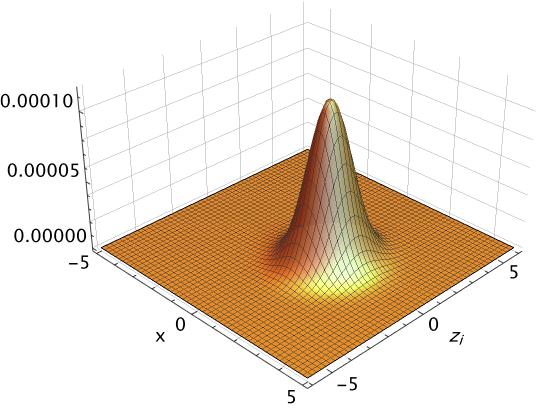}
			\end{subfigure}
			\hfill
			\begin{subfigure}[b]{0.45\textwidth}
				\centering
				\includegraphics[width=\textwidth]{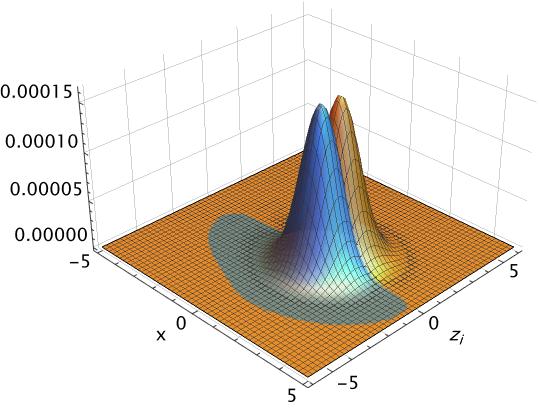}
			\end{subfigure}
			
			\caption{ $\abs{\psi(z;x)}^2$ (in orange) and $\abs{\alpha (z,\lambda)\varphi(z_1;x)}^2$ (in blue) for $m(x)=m_0e^{-x^2}$. 
				On the top we fix $x=1$ and $\lambda=-2$ and we plot the functions for $z_r,z_i$, and $\gamma=1$ (top left) and $\gamma=1+2i$ (top right).
				Bottom: we plot the functions for $x,z_i$ and $z_R=3$, $\lambda_R=-2$, $\gamma=1$ with $\lambda=-2$ (left) and $\lambda=-2+2i$ (right). }
			\label{fig:1}
		\end{figure}

		Substituting this choice of $m(x)$ into equations \eqref{32} and \eqref{33} yields the following bi-coherent states: 
		\begin{equation}
			\varphi(z;x)=\mathcal{M}_{\varphi,z}e^{-z\lambda\sqrt{m_0\pi}\erf\left(\frac{x}{\sqrt{2}}\right)}\varphi_0(x),\qquad \psi(z;x)=\mathcal{M}_{\psi,z}e^{\frac{z\sqrt{m_0\pi}}{\hbar^2}\erf\left(\frac{x}{\sqrt{2}}\right)}\psi_0(x),\label{57}
		\end{equation}
		where $\mathcal{M}_{\varphi,z}$ and $\mathcal{M}_{\psi,z}$ are given in (\ref{310b}) and (\ref{310c}). Some plots of the square modulus of these functions are shown in \autoref{fig:1}. In particular we show two different situations: the self-adjoint cases in the left plots, where the two bicoherent states collapse, and the non self-adjoint cases on the right, where we observe different states. The values of the parameters are given in the caption.

		\subsection{case 2: $m(x)=\frac{m_0}{1+x^2}$}\label{sectV2}
		In this case, the function $m(x)$ tends to zero for large $|x|$ as $1/x^2$. We therefore don't expect square-integrability if the real part of $\lambda$ is positive. The function $F(x)$ is
		\begin{equation}
			\label{58}
			F(x)=\sqrt{m_0}\int_0^x \frac{1}{\sqrt{1+t^2}}dt=\sqrt{m_0}\arcsinh\left(x\right).
		\end{equation}
		The function $\arcsinh(x)$ is monotonically increasing but it is not bounded.  {With this second choice of mass the potential $V(x)$, given by \eqref{27}, is
			\begin{equation}
				V(x)=-\frac{\hbar^2(2+x^2)}{8m_0(1+x^2)}+\frac{\lambda^2 m_0}{2 \hbar^2}\arcsinh^2\left(x\right)-\lambda \gamma \sqrt{m_0}\arcsinh\left(x\right),
			\end{equation}
		}
		and the Hamiltonian can be found as in (\ref{21})
		The two pseudo-bosonic operators became
		\begin{align}
			a &=-\frac{1}{\lambda \sqrt{2m_0}}\left[\sqrt{1+x^2}\left(\dv{}{x}+\frac{x}{2(1+x^2)}\right)-\frac{\lambda m_0}{\hbar^2}\arcsinh(x)  +\gamma \sqrt{m_0}\right], \label{510}  \\
			b &=-\frac{\hbar^2}{\sqrt{2m_0}}\left[\sqrt{1+x^2}\left(\dv{}{x}+\frac{x}{2(1+x^2)}\right)+\frac{\lambda m_0}{\hbar^2}\arcsinh(x)  -\gamma \sqrt{m_0}\right],\label{511}
		\end{align}
		and the ground eigenfunctions 
		\begin{align}
			\varphi_0(x)&=N_{\varphi,0}\left(\frac{m_0}{1+x^2}\right)^{\frac{1}{4}}\exp\left(\frac{\lambda m_0}{2\hbar^2}\arcsinh^2(x)-\gamma \sqrt{m_0}\arcsinh(x)\right),\label{512}\\
			\psi_0(x)&=N_{\psi,0}\left(\frac{m_0}{1+x^2}\right)^{\frac{1}{4}}\exp\left(\frac{\overline{\lambda} m_0}{2\hbar^2}\arcsinh^2(x)-\overline{\gamma}\sqrt{m_0}\arcsinh(x)\right)\label{513}
		\end{align}
		{
			and it follows from (\ref{norm_costant_infinite_case})
			$$N_{\varphi,0} =\overline{N_{\psi,0}}=\left( \hbar \sqrt{\frac{\pi}{-\lambda}} e^{-\frac{\gamma^2 \hbar^2}{\lambda}} \right)^{-\frac{1}{2}}.$$}
		In this case the square-integrability of $	\varphi_0(x)$ and $	\psi_0(x)$ is guaranteed if $\Re(\lambda)<0$.

		Replacing   $m(x)=\frac{m_0}{1+x^2}$ into equations \eqref{32} and \eqref{33} we find 
		\begin{equation}
			\varphi(z;x)=\mathcal{M}_{\varphi,z}e^{-z\lambda\sqrt{2m_0}\arcsinh(x)}\varphi_0(x),\qquad \psi(z;x)=\mathcal{M}_{\psi,z}e^{\frac{z\sqrt{2m_0}}{\hbar^2}\arcsinh(x)}\psi_0(x),\label{515}
		\end{equation}
		where, as usual, $\mathcal{M}_{\varphi,z}$ and $\mathcal{M}_{\psi,z}$ are given in (\ref{310b}) and (\ref{310c}).
		
\begin{figure}[ht!]
			\centering
			\includegraphics[width=0.45\textwidth]{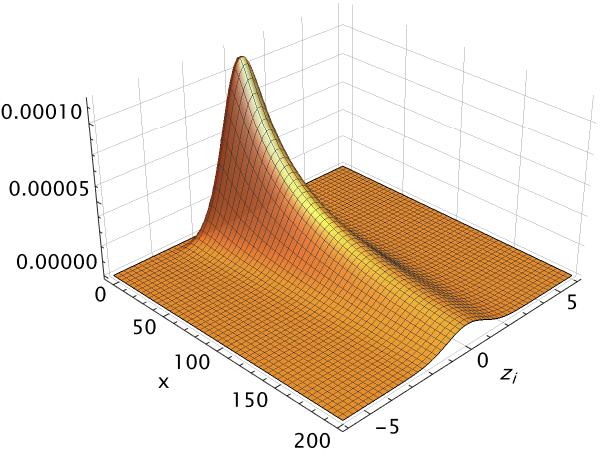}
			\hfill 
			\includegraphics[width=0.45\textwidth]{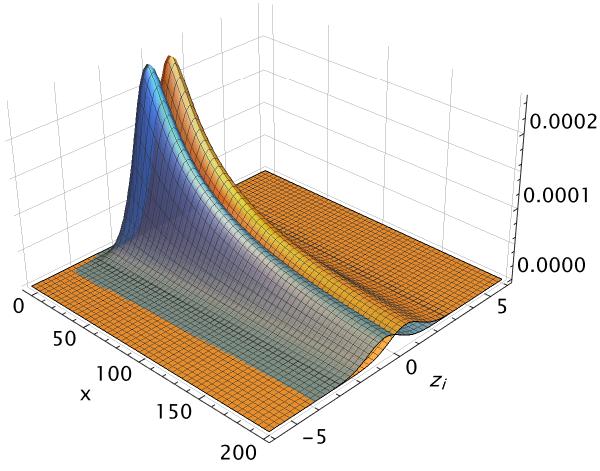}
			
			\caption{ $\abs{\psi(z;x)}^2$ (in orange) and $\abs{\alpha (z,\lambda)\varphi(z_1;x)}^2$ (in blue) for $(x,z_i)$ and $m(x)=\frac{m_0}{1+x^2}$, with $z_r=3$, $\gamma=1$ and $\lambda_R=-2$ and different values of $\lambda_I$. On the left $\lambda_I=0$ ($H$ is self-adjoint). On the  right $\lambda_I=2$  ($H\neq H^\dagger$).}
			\label{fig:2}
		\end{figure}

		Figure \ref{fig:2} depicts the same difference already observed in Figure \ref{fig:1} between self-adjoint and non self-adjoint choice of the parameters.

		\subsection{case 3: $m(x)=m_0\,e^x$}\label{sectV3}

		In this case, the function $m(x)$ tends to zero when $x$ goes to $-\infty$, and diverges when $x\to\infty$. Also, the conditions on the asymptotic behavior of $\frac{1}{m(x)}$ and $\frac{m'(x)}{m^2(x)}$ proposed in Section \ref{sect2} are not satisfied. Still, as we will see, we can use our procedure.  
		
		The function $F(x)$ is
		\begin{equation}
			\label{522}
			F(x)=\sqrt{m_0}\int_0^x e^\frac{t}{2}dt=2\sqrt{m_0e^x},
		\end{equation}
		that is monotonically increasing but not bounded. This third choice of mass produces
			\begin{equation}
				V(x)=-\frac{3\hbar^2e^{-x}}{32m_0}+\frac{2\lambda^2 m_0}{ \hbar^2}e^{x}-2\lambda \gamma \sqrt{m_0e^x},
			\end{equation}
		and the Hamiltonian is given, as usual, by (\ref{21}).
		The two pseudo-bosonic operators became
		\begin{align}
			a &=-\frac{1}{\lambda \sqrt{2m_0e^x}}\left[\dv{}{x}-\frac{1}{4}-\frac{2\lambda m_0}{\hbar^2}e^x  +\gamma \sqrt{m_0e^x}\right], \label{524}  \\
			b &=-\frac{\hbar^2}{\sqrt{2m_0e^x}}\left[\dv{}{x}-\frac{1}{4}+\frac{2\lambda m_0}{\hbar^2}e^x  -\gamma \sqrt{m_0e^x}\right],\label{525}
		\end{align}
		and the ground eigenfunctions 
		\begin{align}
			\varphi_0(x)&=N_{\varphi,0}\left(m_0e^x\right)^{\frac{1}{4}}\exp\left(\frac{2\lambda m_0}{\hbar^2}e^x-\gamma \sqrt{m_0e^x}\right),\label{526}\\
			\psi_0(x)&=N_{\psi,0}\left(m_0e^x\right)^{\frac{1}{4}}\exp\left(\frac{2\overline{\lambda} m_0}{\hbar^2}e^x-\overline{\gamma}\sqrt{m_0e^x}\right),\label{527}
		\end{align}
		{
			where
			\begin{equation}
				N_{\varphi,0}=\overline{N_{\psi,0}}=\left[\frac{i\hbar}{2}\sqrt{\frac{\pi}{\abs{\lambda}}}e^{-\frac{\hbar^2\gamma^2}{\lambda}-i\frac{\theta_{\lambda}}{2}} \left\{1+\erf\left(i\frac{e^{i\frac{\theta_{\lambda}}{2}}\hbar\gamma\sqrt{\abs{\lambda}}}{\lambda}\right) \right\}\right]^{-\frac{1}{2}},
			\end{equation}
			see (\ref{norm_costant_third_case}). Again, the square-integrability of $	\varphi_0(x)$ and $	\psi_0(x)$ is guaranteed if $\Re(\lambda)<0$.
		}
		
		Substituting this choice of $m(x)$ into equations \eqref{32} and \eqref{33} we get: 
		
		\begin{equation}
			\varphi(z;x)=\mathcal{M}_{\varphi,z}e^{-2\sqrt{2}z\lambda\sqrt{m_0e^x}}\varphi_0(x),\qquad \psi(z;x)=\mathcal{M}_{\psi,z}e^{\frac{2\sqrt{2}z\sqrt{m_0e^x}}{\hbar^2}}\psi_0(x)\label{528}
		\end{equation}
		with the standard definition of $\mathcal{M}_{\varphi,z}$ and $\mathcal{M}_{\psi,z}$, see (\ref{310b}) and (\ref{310c}).
		
\begin{figure}[ht!]
			\centering
			\includegraphics[width=0.45\textwidth]{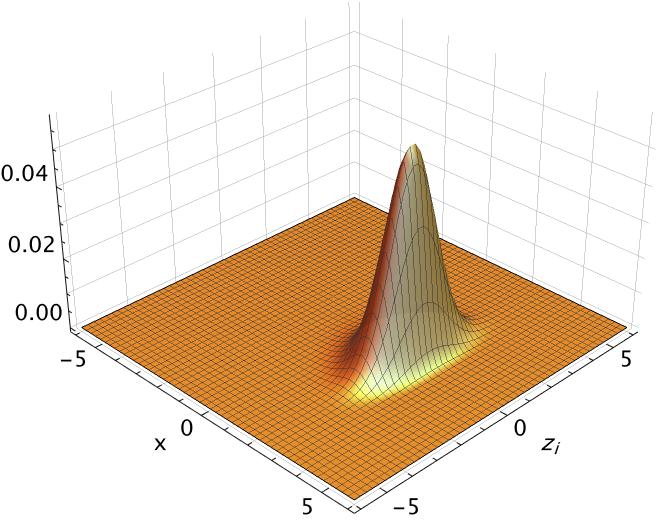}
			\hfill 
			\includegraphics[width=0.45\textwidth]{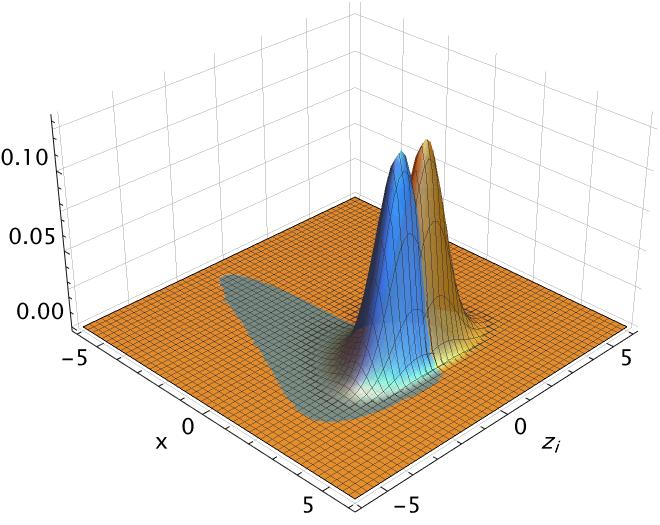}
			
			\caption{$\abs{\psi(z;x)}^2$ (in orange) and $\abs{\alpha (z,\lambda)\varphi(z_1;x)}^2$ (in blue) for $(x,z_i)$ and $m(x)=\frac{m_0}{1+x^2}$, with $z_r=3$, $\gamma=1$ and $\lambda_R=-2$ and different values of $\lambda_I$. On the left $\lambda_I=0$ ($H$ is self-adjoint). On the  right $\lambda_I=2$  ($H\neq H^\dagger$).}
			\label{fig:3}
		\end{figure}

		Figure \ref{fig:3} shows again the same differences beween self-adjoint and non self-adjoint cases for bi-coherent states.
		
		\vspace{2mm}

		{\bf Remark:--} Once again, it is instructive to see what happens in the case of a constant mass: $m(x)=m_0$. We get:
		\begin{equation}
			H=-\frac{\hbar^2}{2m_0}\dv{^2}{x^2}+\frac{\lambda^2 m_0}{2 \hbar^2}x^2-\lambda \gamma \sqrt{m_0}x.\label{516}
		\end{equation}
		It can be observed that the Hamiltonian coincides with the Hamiltonian of the standard or inverted shifted harmonic oscillator, respectively in the cases where $\lambda$ is real or purely immaginary and 
		\begin{equation}
			F(x)=\sqrt{m_0}\int_0^xdt=\sqrt{m_0}x\label{517}.
		\end{equation}
		The two pseudo-bosonic operators became
		\begin{align}
			a &=-\frac{1}{\lambda \sqrt{2m_0}}\left[\dv{}{x}-\frac{\lambda m_0}{\hbar^2}x  +\gamma \sqrt{m_0}\right], \label{518}  \\
			b &=-\frac{\hbar^2}{\sqrt{2m_0}}\left[\dv{}{x}+\frac{\lambda m_0}{\hbar^2}x  -\gamma \sqrt{m_0}\right].\label{519}
		\end{align}
		The vacuum of, e.g., $a$ is
		\begin{equation}
			\varphi_0(x)=N_{\varphi,0} m_0^{\frac{1}{4}}\exp\left(\frac{\lambda m_0}{2\hbar^2}x^2-\gamma \sqrt{m_0}x\right)\label{520}
		\end{equation}
		{	where
			$$N_{\varphi,0} = \left( \hbar \sqrt{\frac{\pi}{-\lambda}} e^{-\frac{\gamma^2 \hbar^2}{\lambda}} \right)^{-\frac{1}{2}}.$$
			This is similar to case 2 above, since in both cases the function \(F(x)\) is not bounded for $|x|$ diverging.}
		
		As we expect, if $\lambda=-1, \gamma=0$, we obtain the eigenstate of the harmonic oscillator. If $\lambda=\pm i, \gamma=0$, we obtain the eigenstate of the inverted harmonic oscillator, and we lose the property of having square-integrable functions\footnote{In this case, of course, $N_{\varphi,0}$ is not really a normalization constant. Still, we use this name instead of calling $N_{\varphi,0}$ an "integration constant".}.

		Substituting this choice of $m(x)$ into equations \eqref{32} and \eqref{33} yields the following bi-coherent states: 
		\begin{equation}
			\label{521}
			\varphi(z;x)=\mathcal{M}_{\varphi,z}e^{-z\lambda\sqrt{2}\sqrt{m_0}x}\varphi_0(x),\qquad \psi(z;x)={M}_{\psi,z}e^{\frac{z\sqrt{2}}{\hbar^2}\sqrt{m_0}x}\psi_0(x).
		\end{equation}
		with the usual definitions of $\mathcal{M}_{\varphi,z}$ and $\mathcal{M}_{\psi,z}$.

		\section{Conclusions}\label{sect6}
		
		In this paper, we have investigated a one-dimensional Hamiltonian characterized by a position dependent mass and in presence of a complex potential, using the general procedure proposed for abstract ladder operators. By requiring the commutation relation $\comm{H}{A} = \lambda A$, together with the factorizability of a shifted version of $H$, we have derived the explicit functional forms of the operators $A$ and $B$ used to factorize $H$, as well as the expression of the complex potential $V(x)$, in terms of the mass $m(x)$ and of the (in general) free parameter $\lambda$.
		
		Our results demonstrate that the pseudo-bosonic algebraic structure is naturally preserved even in presence of a non-constant mass, provided that the potential is suitably adapted. We have constructed two families of biorthogonal eigenstates for $H$ and $H^\dagger$ in terms of Hermite polynomials of a function of the mass. 
		
		In Section \ref{sect4} we have shown that the very general expression of the potential (\ref{27}) is indeed connected, with a suitable transformation, to some quadratic potential as in (\ref{45}). This is in agreement with the fact that the eigenvalues of $H$ in (\ref{21}) are, with our choices, always linear in $n$, see (\ref{224}).

		Furthermore, we have introduced bi-coherent states using two different approaches: as eigenstates of the lowering operators and as the result of the action of displacement-like operators on the ground states, or (equivalently) as suitable convergent power series. The equivalence of these two definitions has been proven by means of the generating function for Hermite polynomials.
		
		We have also considered few different specific choices for $m(x)$, which produce different situations and localization characteristics.

		\section*{Acknowledgments}  F. B. acknowledges partial support from Palermo University and from G.N.F.M. of the INdAM. F, B. also  acknowledges partial support from PRIN Project {\em  Transport phonema in low dimensional structures: models, simulations and theoretical aspects}, and partial support  by  the project ICON-Q, Partenariato Esteso NQSTI - PE00000023, Spoke 2.

		\renewcommand{\theequation}{A.\arabic{equation}}
		
		\section*{Appendix A: Some aspects of ALOs}\label{appendix}
		
		Suppose that the Hamiltonian $H$, which is not necessarily self-adjoint, $H\neq H^\dagger$, admits some ladder operator:
		\be
		[H,Z_j]=\lambda_j Z_j,
		\label{a1}\en
		$\lambda_j\in\mathbb{C}$, for $j=1,2,\ldots,N$. This is equivalent to require that
		\be
		[H^\dagger,Z_j^\dagger]=-\overline{\lambda_j}\, Z_j^\dagger,
		\label{a2}\en
		or that
		\be
		[H,Z_j^n]=n\lambda_j Z_j^n, \qquad [H^\dagger,{Z_j^\dagger}^n]=-n\overline{\lambda_j}\, {Z_j^\dagger}^n, 
		\label{a3}\en
		for all $n=0,1,2,\ldots$ and $j=1,2,\ldots,N$. 
		
		Let us suppose that two, in general different, nonzero vectors exist, $\varphi_E$ and $\psi_E$, such that
		\be
		H\varphi_E=E\varphi_E, \qquad H^\dagger\psi_E=\overline{E}\,\psi_E.
		\label{a4}\en
		These are respectively eigenstates of $H$ and $H^\dagger$, with complex conjugate eigenvalues. We define the following vectors
		\be
		\varphi_{E:j,n}=Z_j^n\varphi_E, \qquad \psi_{E:j,n}={Z_j^\dagger}^n\psi_E,
		\label{a5}\en
		for all $n=0,1,2,\ldots$ and $j=1,2,\ldots,N$. Of course, if some integer $n_0(j)$ exists such that $\varphi_{E:j,n_0(j)}=0$, it follows that $\varphi_{E:j,n}=0$ for all $n\geq n_0(j)$. Similarly, if some integer $m_0(j)$ exists such that $\psi_{E:j,m_0(j)}=0$,  it follows that $\psi_{E:j,m}=0$ for all $m\geq m_0(j)$. It is easy to show that each non zero $\varphi_{E:j,n_0(j)}$ is an eigenstate of $H$. Analogously, each non zero $\psi_{E:j,n}$ is an eigenstate of $H^\dagger$, \cite{bagalo1}: 
		\be
		H\varphi_{E:j,n}=\epsilon_{E:j,n}\varphi_{E:j,n}, \qquad H^\dagger\psi_{E:j,n}=\overline{\epsilon_{E:j,-n}}\,\psi_{E:j,n},
		\label{a6}\en
		where
		\be
		\epsilon_{E:j,n}=E+n\lambda_j.
		\label{a7}\en
		Calling $\F_\varphi=\{\varphi_{E:j,n}:\, j=1,2,\ldots, N; \,n\geq0\}$ and $\F_\psi=\{\psi_{E:j,n}:\, j=1,2,\ldots, N; \,n\geq0\}$, these two sets are biorthogonal in the following sense:
		\be
		\mbox{if } n\lambda_j\neq -m\lambda_l \quad\Rightarrow\quad\langle\psi_{E:l,m},\varphi_{E:j,n}\rangle=0.
		\label{a8}
		\en
		A lot of other results and examples on ALOs can be found, in particular, in  \cite{bagalo1,bagalo2}.
		
		\setcounter{equation}{0}
		
		\renewcommand{\theequation}{B.\arabic{equation}}

		\section*{Appendix B: Derivation of eigenstates}
		
		In this section we discuss how the functions $\varphi_n(x)$ in (\ref{219}) can be deduced.
		We can easily compute the first non normalized excited state  $\tilde{\varphi}_1(x)=b\varphi_0(x)$
		\begin{equation}
			\tilde{\varphi}_1(x)=b\varphi_0(x)=\frac{2}{\sqrt{2}}\left(\hbar^2\gamma-\lambda F(x)\right)\varphi_0(x)=2\Gamma(x)\varphi_0(x),\label{a14}
		\end{equation}
		where we introduced the function $\Gamma(x)=\frac{1}{\sqrt{2}}\left(\hbar^2\gamma-\lambda F(x)\right)$.
		
		In order to simplify the calculations, it is convenient to decompose the operator $b$ as $b=C+D$, where
		\begin{equation}
			C=-\frac{\hbar^2}{\sqrt{2m(x)}}\dv{}{x}, \qquad D=\frac{\hbar^2}{\sqrt{2}}\left(\frac{m^{-3/2}(x)m'(x)}{4}+\gamma-\frac{\lambda}{\hbar^2}F(x)\right).\label{a15}
		\end{equation}
		In this form, we can apply the Leibniz rule to the operator $C$. Let us compute $b(\Gamma(x)\tilde{\varphi}_n)$:
		\begin{align*}
			b(\Gamma(x)\tilde{\varphi}_n)&=C(\Gamma(x)\tilde{\varphi}_n)+D(\Gamma(x)\tilde{\varphi}_n)= \\
			&=C(\Gamma(x))\tilde{\varphi}_n+\Gamma(x)C(\tilde{\varphi}_n)+\Gamma(x)D(\tilde{\varphi}_n)= \\
			&=C(\Gamma(x))\tilde{\varphi}_n+\Gamma(x)(C+D)(\tilde{\varphi}_n)= \\
			&=C(\Gamma(x))\tilde{\varphi}_n+\Gamma(x)b(\tilde{\varphi}_n).
		\end{align*}
		Let us define the complex constant $k$. Using the polar representation for $\lambda$, $\lambda=|\lambda|e^{i\theta_\lambda}$, we have
		\begin{equation}
			k = -\frac{\lambda\hbar^2}{2}.\label{a16}
		\end{equation}
		so that
		\begin{equation}
			\sqrt{k} = \frac{\hbar\sqrt{\abs{\lambda}}}{\sqrt{2}}e^{\frac{i}{2}\left(\theta_\lambda  + \pi\right)}.\label{a17}
		\end{equation}
		Noting that $C(\Gamma(x)) = -k$, and recalling that $b\tilde{\varphi}_n=\tilde{\varphi}_{n+1}$, we obtain the recursive relation:
		\begin{equation}
			b(\Gamma(x)\tilde{\varphi}_n)=\Gamma(x)\tilde{\varphi}_{n+1}-k\tilde{\varphi}_{n}.\label{a18}
		\end{equation}
		From this result, since $\tilde{\varphi}_1 = 2\Gamma \varphi_0 = 2\Gamma \tilde{\varphi}_0$, we immediately find that $\tilde{\varphi}_2=2\Gamma(x)\tilde{\varphi}_1-2k\tilde{\varphi}_0$. It is straightforward to prove by induction that
		\begin{equation}
			\tilde{\varphi}_{n+1}=2\Gamma(x)\tilde{\varphi}_{n}-2nk\tilde{\varphi}_{n-1}.\label{a19}
		\end{equation}
		This recurrence relation implies that $\tilde{\varphi}_n(x)$ can be written as $\tilde{\varphi}_n(x)=P_n(\Gamma(x))\varphi_0(x)$, where $P_n$ is a polynomial of degree $n$ satisfying $P_{n+1}(\Gamma)=2\Gamma P_{n}(\Gamma)-2nk P_{n-1}(\Gamma)$.
		
		By direct computation of the first few terms:
		\begin{align*}
			P_0(\Gamma)&=1= H_0\left(\frac{\Gamma}{\sqrt{k}}\right), \\
			P_1(\Gamma)&=2\Gamma=2\frac{\Gamma}{\sqrt{k}}\sqrt{k}=\sqrt{k} H_1\left(\frac{\Gamma}{\sqrt{k}}\right), \\
			P_2(\Gamma)&=2(2\Gamma^2-k)=2\left(2\left(\frac{\Gamma}{\sqrt{k}}\right)^2-1\right)k=k H_2\left(\frac{\Gamma}{\sqrt{k}}\right),
		\end{align*}
		where $H_n$ denotes the $n$-th Hermite polynomial. We assume the general form of the polynomial is $P_n(\Gamma)=(\sqrt{k})^nH_n\left(\frac{\Gamma}{\sqrt{k}}\right)$. Substituting this ansatz into the recurrence relation and dividing by $(\sqrt{k})^n$:
		\begin{align*}
			\frac{1}{\sqrt{k}}P_{n+1}(\Gamma)&=2\frac{\Gamma}{\sqrt{k}}P_{n}(\Gamma)-2n\sqrt{k}P_{n-1}(\Gamma) \\
			\frac{1}{\sqrt{k}}(\sqrt{k})^{n+1}H_{n+1}\left(\frac{\Gamma}{\sqrt{k}}\right)&=2\frac{\Gamma}{\sqrt{k}}\left(\sqrt{k}\right)^{n}H_{n}\left(\frac{\Gamma}{\sqrt{k}}\right)-2n\sqrt{k}\left(\sqrt{k}\right)^{n-1}H_{n-1}\left(\frac{\Gamma}{\sqrt{k}}\right) \\
			H_{n+1}\left(\frac{\Gamma}{\sqrt{k}}\right)&=2\frac{\Gamma}{\sqrt{k}}H_{n}\left(\frac{\Gamma}{\sqrt{k}}\right)-2nH_{n-1}\left(\frac{\Gamma}{\sqrt{k}}\right).
		\end{align*}
		This matches the standard recurrence relation for Hermite polynomials, proving the ansatz. Finally, the normalized eigenstates $\varphi_n(x)$ are given by:
		\begin{equation}
			\begin{split}
				\varphi_n(x)&=\frac{1}{\sqrt{n!}}\tilde{\varphi}_n(x)=\frac{1}{\sqrt{n!}}P_n(\Gamma(x))\varphi_0(x)= \\
				&=\frac{1}{\sqrt{n!}}\left(\frac{\hbar\sqrt{\abs{\lambda}}}{\sqrt{2}}e^{\frac{i}{2}\left(\theta_\lambda  + \pi\right)}\right)^{n}H_n\left(\frac{\hbar^2\gamma-\lambda F(x)}{\hbar\sqrt{\abs{\lambda}}}e^{-\frac{i}{2}\left(\theta_\lambda+\pi\right)}\right)\varphi_0(x),\label{a20}
			\end{split}
		\end{equation}
		where we used the explicit polar form for the square root in the coefficient and in the Hermite argument.
		
		In a similar way one can prove the expression (\ref{220}) for the $\psi_n(x)$.

		\setcounter{equation}{0}
		
		\renewcommand{\theequation}{C.\arabic{equation}}

		\section*{Appendix C: The normalization of $\varphi_0(x)$ and $\psi_0(x)$}\label{appendixc}

		Requiring that $\langle\psi_0,\varphi_0\rangle=1$ and that $N_{\varphi,0}=\overline{N_{\psi,0}}$ we find that 
		\begin{equation}
			N_{\varphi,0}=\overline{N_{\psi,0}}=\left[\int_\R\sqrt{m(x)}\exp\left\{\frac{\lambda}{\hbar^2}F^2(x)-2\gamma F(x)\right\}dx\right]^{-\frac{1}{2}}
		\end{equation}
		With the same change of variables as in Section \ref{sectII2}, it follows that
		\begin{equation}
			N_{\varphi,0}=\overline{N_{\psi,0}}=\left[\int_{F_{-\infty}}^{F_{\infty}}\exp\left\{\frac{\lambda}{\hbar^2}F^2-2\gamma F\right\}dF\right]^{-\frac{1}{2}},
			\label{integral_norm_costant}
		\end{equation}
		this shows that the normalization constant does not depend on the entire mass function, but rather by the boundedness of the function $F(x)$.
		The integral (\ref{integral_norm_costant}) can be solved using the complex error function
		\begin{equation}
			\erf(z)=\frac{2}{\sqrt{\pi}}\int_0^{z}e^{-t^2}dt
		\end{equation}
		and it follows that
		\begin{equation}
			N_{\varphi,0}=\left\{\frac{i\hbar}{2}\sqrt{\frac{\pi}{\abs{\lambda}}}e^{-\frac{\hbar^2\gamma^2}{\lambda}-i\frac{\theta_{\lambda}}{2}} \left[\erf\left(i\frac{e^{i\frac{\theta_{\lambda}}{2}}\sqrt{\abs{\lambda}}}{\hbar}\left\{F_{\infty}-\frac{ \hbar^2\gamma}{\lambda}\right\}\right)-\erf\left(i\frac{e^{i\frac{\theta_{\lambda}}{2}}\sqrt{\abs{\lambda}}}{\hbar}\left\{F_{-\infty}-\frac{ \hbar^2\gamma}{\lambda}\right\}\right) \right]\right\}^{-\frac{1}{2}},
			\label{norm_constant_general_case}
		\end{equation}
		where, for concreteness , we fix $\sqrt{i}=e^{i\frac{\pi}{4}}$.
		
		In particular, if $F_{\infty}=-F_{-\infty}=\infty$, formula (\ref{norm_constant_general_case}) becomes
		\begin{equation}
			N_{\varphi,0}=\overline{N_{\psi,0}}=\left(i\hbar\sqrt{\frac{\pi}{\abs{\lambda}}}e^{-\frac{\hbar^2\gamma^2}{\lambda}-i\frac{\theta_{\lambda}}{2}}\right)^{-\frac{1}{2}}.
			\label{norm_costant_infinite_case}
		\end{equation}
		If we rather have a finite $F_{-\infty}$, with $F_{\infty}$ infinite, it follows that  
		\begin{equation}
			N_{\varphi,0}=\overline{N_{\psi,0}}=\left\{\frac{i\hbar}{2}\sqrt{\frac{\pi}{\abs{\lambda}}}e^{-\frac{\hbar^2\gamma^2}{\lambda}-i\frac{\theta_{\lambda}}{2}} \left[1-\erf\left(i\frac{e^{i\frac{\theta_{\lambda}}{2}}\sqrt{\abs{\lambda}}}{\hbar}\left\{F_{-\infty}-\frac{ \hbar^2\gamma}{\lambda}\right\}\right) \right]\right\}^{-\frac{1}{2}},
			\label{norm_costant_third_case}
		\end{equation}
		while a similar expression can be deduced if $F_{-\infty}$ is infinite and $F_{\infty}$ is finite.


\begin{thebibliography}{99}
			
			\bibitem{benboet} C. M. Bender, S. Boettcher, {\em Real Spectra in Non-Hermitian Hamiltonians Having PT-Symmetry}, Phys. Rev. Lett.
			{\bf 80}, 5243-5246, (1998)
			
			
			\bibitem{specissue2012} C. Bender, A. Fring, U. G\"{u}enther, H. Jones Eds, {\em Special issue on quantum physics with non-Hermitian operators}, J. Phys. A: Math. and Ther., {\bf 45} (2012)
			
			\bibitem{bagbook2015} F. Bagarello, J. P. Gazeau, F. H. Szafraniec e M. Znojil Eds., {\em Non-selfadjoint operators in quantum physics: Mathematical aspects}, John Wiley and Sons (2015)
			
			
			
			\bibitem{benbook} C. M. Bender,  {\em $PT$ Symmetry In Quantum and Classical Physics}, World Scientific Publishing Europe Ltd., London (2019)
			
			
			\bibitem{bagabook} F. Bagarello, J. P. Gazeau, F. H. Szafraniec e M. Znojil Eds., {\em Non-selfadjoint operators in quantum physics: Mathematical aspects}, John Wiley and Sons (2015)
			
			\bibitem{bagprocpa} F. Bagarello, R. Passante, C. Trapani, {\em Non-Hermitian Hamiltonians in Quantum Physics;
				Selected Contributions from the 15th International Conference on Non-Hermitian
				Hamiltonians in Quantum Physics}, Palermo, Italy, 18-23 May 2015, Springer (2016)
			
			
			
			
			\bibitem{bagspringerbook} F. Bagarello, {\em Pseudo-Bosons and Their Coherent States}, Springer, Mathematical Physics Studies, 2022
			
			
			\bibitem{mosta} A. Mostafazadeh, {\em Pseudo-hermitian quantum mechanics},  Int. J. Geom. Methods Mod. Phys., {\bf 7}, 1191-1306 (2010)
			
			\bibitem{pdmass1}
			O. von Roos, {\em Position-dependent effective masses in semiconductor theory}, Phys. Rev. B, {\bf 27}, 7547–7552 (1983)
			
			
			\bibitem{pdmass2} B. Bagchi, A. Banerjee, C. Quesne, V. M. Tkachuk,  {\em Deformed shape invariance and exactly solvable
				Hamiltonians with position-dependent effective mass}, J. Phys. A: Math. Gen., {\bf 38}, 2929–2945 (2005)
			
			\bibitem{pdmass3} S. Cruz y Cruz, O. Rosas-Ortiz, {\em Position-dependent mass oscillators and coherent states}, J. Phys. A: Math.
			Theor., {\bf 42}, 185205 (2009)
			
			\bibitem{estrada} M. I. Estrada-Delgado, Z. I. Blanco-Garcia, {\em Oscillator Algebra in Complex Position-Dependent Mass Systems},  Symmetry, {\bf 17}, 10, 1690 (2025)	
			
			\bibitem{takou} D. S. Takou, A. Y. Mora, I. Nonkan\'e,  L. M. Lawson,  G. Y. H. Avossevou,  {\em Gazeau-Klauder coherent states for a harmonic position-dependent mass}, J. Comput. Electron., {\bf 24}, 191 (2025)
			
			\bibitem{quesne} C. Quesne, {\em Semi-infinite quantum wells in a position-dependent mass background}, Quantum Stud.: Math. Found. {\bf 10} 237-244, (2023)
			
			
			
			
			\bibitem{art2008a} O. Mustafa, S. H. Mazharimousavi, {\em Complexified von Roos Hamiltonian's $\eta$-weak-pseudo-Hermiticity, isospectrality and exact solvability}, J. Phys. A: Math. Theor. {\bf 41} 244020 (2008) 
			
			\bibitem{art2008b} O. Mustafa, S. H. Mazharimousavi,  {\em First-Order Intertwining Operators with Position Dependent Mass and $\eta$-Weak-Pseudo-Hermiticity Generators}, Int J Theor Phys {\bf 47}, 446–454 (2008)
			
			\bibitem{art2025} A. Ballesteros, R. Ramírez, M. Reboiro, {\em Non-standard quantum algebras and infinite-dimensional PT-symmetric systems}
			J. Phys. A: Math. Theor. {\bf 58}, 455301 (2025)
			
			
			
			
			
			\bibitem{art2023} F. C. E. Lima, L. N. Monteiro, C. A. S. Almeida,
			{\em Non-Hermitian fermions with effective mass},
			Physica E: Low-dimensional Systems and Nanostructures,
			{\bf 150}, 115682 (2023)
			
			\bibitem{art2006} C.-S. Jia, A de Souza Dutra,{\em Position-dependent effective mass Dirac equations
				with PT-symmetric and non-PT-symmetric potentials}, J. Phys. A:
			Math. Gen. {\bf 39}, 11877–11887 (2006) 
			
			
			
			
			
			
			
			\bibitem{fern1} F. M. Fernandez, {\em Algebraic treatment of PT-symmetric coupled oscillators}, Int. J. Theor. Phys. {\bf 54} 3871-3876 (2015)
			
			\bibitem{fern2} F. M. Fernandez, {\em Symmetric quadratic Hamiltonians with pseudo-Hermitian matrix representation}, Ann. Phys. {\bf 369}, 168-176 (2016)
			
			\bibitem{fern3}  F. M. Fernandez, {\em Algebraic treatment of non-Hermitian quadratic Hamiltonians}, J. Math. Chem., {\bf 58}, 2094--2107 (2020)
			
			
			\bibitem{bagalo1} F. Bagarello,  {\em
				Abstract ladder operators and their applications},  J. Phys. A,  {\bf 54}, 445203 (2021)
			
			
			\bibitem{bagalo2} F. Bagarello,  {\em  Abstract ladder operators for non self-adjoint Hamiltonians, with applications},   Ann. of Phys.,  {\bf 468}, 169727
			(2024)	
			
			
			\bibitem{reed} S. Reed, B. Simon, {\em Methods of modern mathematical physics}, Vol I: {\em Functional analysis}, Academic, New York (1975)
			
			
			
			
			\bibitem{barton} G. Barton, {\em Quantum mechanics of the inverted oscillator potential},  Ann. of Phys., {\bf 166}, 322-363 (1986) 
			
			
			\bibitem{krason} P. Krason, J. Milewski, {\em On eigenproblem for inverted harmonic oscillators}, Banach Center Publications, {\bf 124}, 61-73  (2021)
			
			\bibitem{subra} V. Subramanyan, S. S. Hegde, S. Vishveshwara, B. Bradlyn, {\em Physics of the inverted harmonic oscillator: from the lowest Landau level to event 	horizons}, Ann. of Phys., {\bf 435}, 168470 (2021) 
			
			\bibitem{sunda} Sriram Sundaram, C. P. Burgess, D. H. J. O’Dell, {\em Duality between the quantum inverted harmonic oscillator and
				inverse square potentials}, New J. Phys. {\bf 26} (2024) 053023
			
			\bibitem{fujita} T. Fujita, Y. Kaku, A. Matsumura, Y. Michimura, {\em Inverted oscillators for testing gravity-induced quantum entanglement}, Class. Quantum Grav. {\bf 42}, 165003 (2025)
			
			\bibitem{bagiqho} F. Bagarello,  {\em A Swanson-like Hamiltonian and the inverted harmonic oscillator},  J. Phys. A, {\bf 55}. 225204 (2022)
			
			
			\bibitem{bagbaku} F. Bagarello, E. Balistreri, S.Ku\.{z}el, {\em
				An unified approach for the (inverted) quantum harmonic oscillator and the Berry-Keating hamiltonian},  J. Math. Phys., submitted
			
			\bibitem{bag2010} F. Bagarello, {\em Examples of Pseudo-bosons in quantum mechanics},  Phys. Lett. A,  {\bf 374}, 3823-3827 (2010)
			
			
			\bibitem{baginbagbook} F. Bagarello, {\em Deformed canonical (anti-)commutation relations and non hermitian Hamiltonians}, in {Non-selfadjoint operators in quantum physics: Mathematical aspects}, F. Bagarello, J. P. Gazeau, F. H. Szafraniec and M. Znojil Eds., Wiley  (2015)
			
			\bibitem{kolfom} A. Kolmogorov and S. Fomine, {\em El\'ements de la th\'eorie des fonctions et de l'analyse fonctionnelle}, Mir (1973)
			
			
			
			
			
			
			
			
			
			
			
			
			
			
			
			
			
			
			
			
			
			
			%
			%
			%
			%
			%
			%
			%
			%
			%
			%
			%
			%
			%
			%
			%
			%
			%
			%
			
			
		\end{thebibliography}
	\end{document}